\documentclass[twocolumn,prb,showpacs,floatfix,amsfonts]{revtex4}
\pdfoutput=1

\usepackage[latin9]{inputenc}
\usepackage{textcomp}
\usepackage{amsmath}
\usepackage{amssymb}
\usepackage{graphicx}
\usepackage[pdftex]{hyperref}

\makeatletter

\newcommand{\beq}{\begin{equation}}
\newcommand{\eeq}{\end{equation}}

\makeatother

\begin{document}

\title{Renormalization of nanoparticle polarizability in the vicinity of a graphene-covered interface}

\author{Jaime E. Santos}\email{jaime.santos@fisica.uminho.pt} \author{M. I. Vasilevskiy} \author{N. M. R. Peres} \author{G. Smirnov} \author{Yu. V. Bludov }

\affiliation{Centro de F\'{i}sica and Departamento de F\'{i}sica,
Universidade do Minho, P-4710-057 Braga, Portugal}

\begin{abstract}
We study the electromagnetic properties of a metamaterial consisting of polarizable (nano)particles and a single graphene sheet placed at the interface between two dielectrics. We show that the particle's polarizability is renormalized because of the electromagnetic coupling to surface plasmons supported by graphene, which results in a dispersive behavior, different for the polarizability components corresponding to the induced dipole moment, parallel and perpendicular to the graphene sheet.
In particular, this effect is predicted to take place for a metallic particle whose bare polarizability in the terahertz (THz) region is practically equal to the cube of its radius (times $4\pi \varepsilon _0$).
This opens the possibility to excite surface plasmons in graphene and enhance its absorption in the THz range by simply using a monolayer of metallic particles randomly deposited on top of it, as we show by explicit calculations.

\end{abstract}


\maketitle

\section{Introduction}

Electromagnetic (EM) metamaterials are artificial structures designed in such a way that their optical properties differ from those existing in natural
materials.\cite{Engheta2006,Shalaev2011} They offer new functionalities, such as radiation guiding\cite{Bozhevolnyi2013}, enhanced absorption and EM energy concentration in sub-wavelength regions,\cite{Kravets2008,Ferry} extraordinary  transmission,\cite{GarciadeAbajo2007} color filtering\cite{TingXu2010} and tailoring\cite{Torell}, surface-enhanced Raman scattering (SERS),\cite{Kim} {\it etc.} Many of these unusual properties are related to surface plasmons, collective oscillations of free electrons, which either propagate along a conductor's surface or a nanowire, or are localized in a metallic nanoparticle (NP).\cite{Novotny-Hecht,Blackman}
Graphene, a two-dimensional conductor, possesses unusual electronic properties\cite{rmp}, and graphene plasmonics\cite{novnatphoton} has become a field of intense research, both theoretical and experimental; see Refs. \onlinecite {c:primer,Luo-r2013} for reviews.
It offers the possibility of expansion of metamaterials to the far-infrared (FIR) and THz spectral range and allows for their tunability, most directly achieved by adjusting the Fermi level in graphene through an external gate voltage\cite{Li2008}, but also in a number of different ways, which can be implemented by using periodic structures of graphene ribbons,\cite{Ju2011,nikitin_ribbon2012}
two-\cite{Yan_disks2012,Thongrattanasiri2012,Fang2014} or three-dimensional\cite{Berman2010,Yan2012} arrays of graphene disks, or a two-dimensional array of antidots.\cite{nikitin2012}

A potentially interesting direction of research is combining graphene with quasi-zero-dimensional emitters or absorbers, such as organic molecules\cite{Gaudreau2013,Agarwal2013,Velizhanin2011} or semiconductor quantum dots (QDs) \cite{Chen2010,Konstantatos2012}. Such study explores the possibilities of electromagnetic coupling between localized excitations (for instance, molecular or QD excitons) and propagating graphene plasmons in order to probe the de-excitation dynamics\cite{Gaudreau2013} or dispersion relation of plasmon-polaritons in graphene \cite{Velizhanin2011}, enhance the F\"orster transfer between an emitter and an absorber,\cite{Agarwal2013} control the coupling between two emitters (superradiance effect)\cite{Huidobro2012}, or enhance the EM radiation absorption in graphene.\cite{Stauber2014}
Another possibility that has been recently demonstrated experimentally\cite{Chen2012,Fei2012}
is that of electromagnetic coupling between the said graphene plasmons and an illuminated atomic force microscope tip,
which allows for the study of the plasmon dispersion relation as a function of the gating applied to graphene.

Qualitatively similar effects have been predicted and observed, in the visible range, for hybrid systems with metal plasmons; for instance, generation of single optical plasmons in metallic nanowires coupled to QDs\cite{Akimov2007}, metal-enhanced\cite{Ray2013} or quenched\cite{Lunz2012,Schreiber2014} fluorescence of colloidal semiconductor nanocrystals, or resonant absorption by exciton-plasmon polaritons.\cite{Gomez2010,Bludov2012} However, the case of graphene is special not only because
it involves a different spectral region, but also because graphene is a semimetal and its plasma oscillations are mediated by both intraband and interband transitions, with a characteristic frequency-dependent conductivity.\cite{rmp} Moreover, since it is a monolayer-thick material, it should be considered as a two-dimensional (2D) object rather than a very thin 3D film.\cite {c:primer} As a result, graphene, for instance, supports both $p$- and $s$-polarized surface waves.\cite{Mikhailov2007} In such a case, the EM coupling to non-plasmonic excitations may also have features that are not known for metal surface plasmons.

The aim of the present paper is to provide both a qualitative and quantitative account of the electromagnetic properties of a metamaterial
consisting of polarizable (nano)particles and a single graphene sheet placed at the interface betwen two dielectrics, one of which incorporates the particles.
Using the electrostatic approximation, we calculate the field created by polarization charges induced on the graphene sheet by the particle excited by an external EM field, as well as the said surface charge density on graphene, and describe the resulting effect in terms of its renormalized polarizability. The calculation of the frequency-dependent renormalized polarizability is the
main result of this article. We show that it is a second-rank tensor with two unequal principal values, which can have a pronounced dependence upon the excitation frequency even if particle's polarizability is nondispersive in the considered THz spectral range. In particular, this effect is shown to take place for a spherical gold particle lying on the graphene sheet. Once the renormalized polarizability of a single particle is computed, the EM properties (i.e., reflection, transmission and absorption spectra) of the metamaterial consisting of a graphene layer sandwiched between two dielectrics, one of which is doped with polarizable particles, can be calculated.
We explicitly compute the THz optical properties of a monolayer of nonabsorbing nanoparticles randomly deposited on top of a graphene sheet and show that the absorption in graphene is enhanced due to the excitation of surface plasmons.

This article is organized as follows. In Sec. \ref{secA}, we define our model system and derive the electrostatic boundary conditions on graphene. The electric fields are obtained using the method of images in Sec. \ref{secB}. In  Sec. \ref{secC}, the renormalized polarizability is introduced and a few examples, involving nanoparticles  constituted of different materials, are also discussed. The following two sections present the calculated results for the polarization charge density induced on graphene and the THz optical spectra of the system composed of a monolayer of polarizable particles randomly deposited on the graphene sheet. We conclude in Sec \ref{sec_conclusion}.

\section{Model system and associated boundary conditions}
\label{secA}
\begin{figure}
\begin{center}
\includegraphics[width=7.5cm]{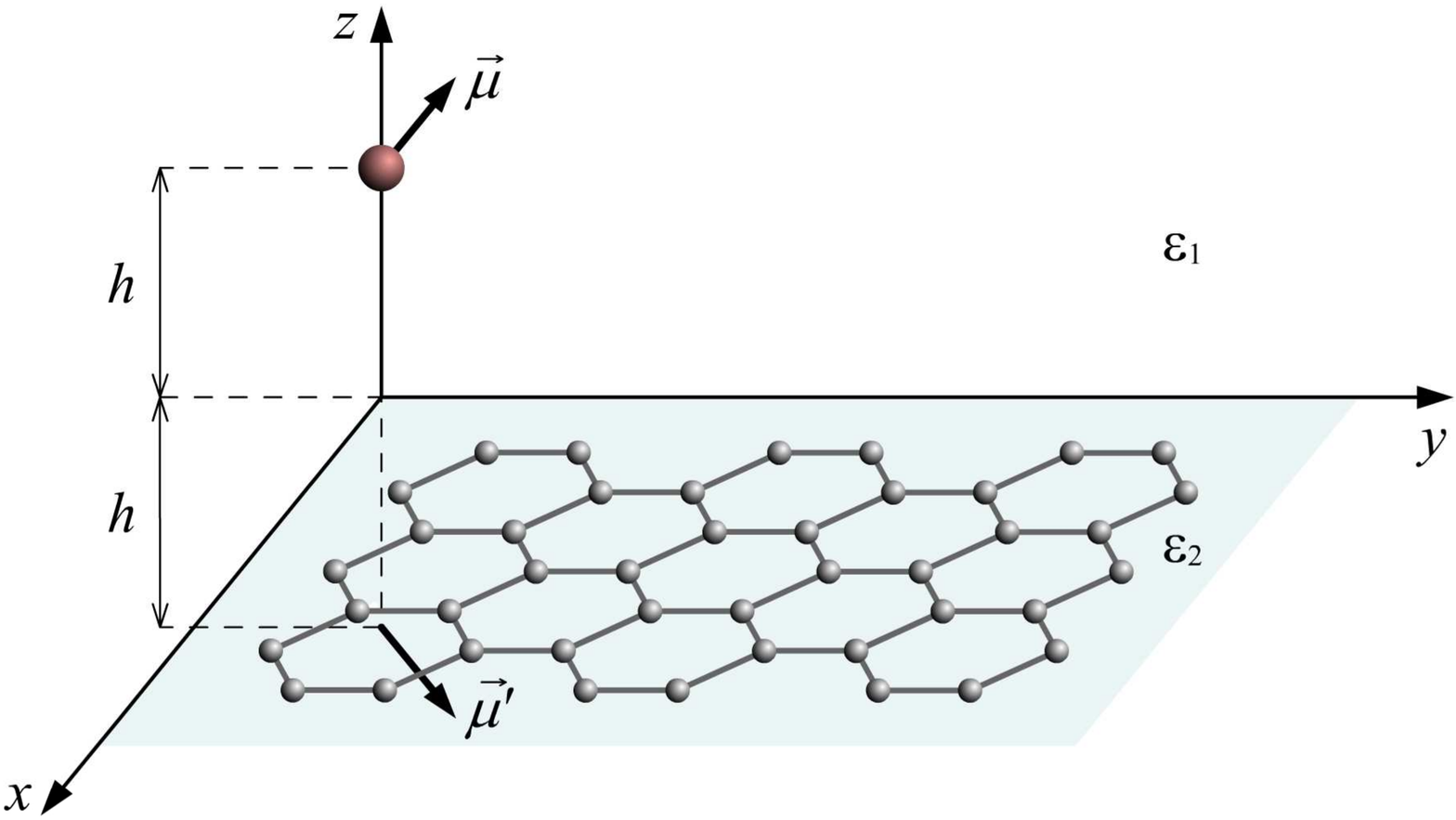}
\end{center}
\caption{(Color online) Schematics of the system consisting of a dipole (polarized NP) located at a point $(0,\!0,\!h)$ in the vicinity of a graphene-covered interface between two dielectrics. An image dipole located at $(0,\!0,\!-h)$ is also shown.}
\label{fig1}
\end{figure}

We first consider the problem of a single polarizable (nano)particle,
placed in the vicinity of the interface between two dielectrics, with relative permittivities
that can depend on the frequency $\omega$ of the electric field in the media,
given by $\varepsilon_1(\omega)$
(in medium 1, the upper medium) and $\varepsilon_2(\omega)$
(in medium 2, the lower medium). The particle is located in medium 1 at a distance
$h$ along the normal to the interface, the interface
being identified with the plane $z=0$ in our system of coordinates [the position
of the particle is given by $\mathbf{r}_0=(0,0,h)$; see Fig. \ref{fig1}]. The interface
is constituted by a graphene
sheet, which is homogeneous in a macroscopic scale and is described at this scale by its
(frequency-dependent) conductivity $\sigma(\omega)$.
We will treat the problem in the electrostatic approximation where one can neglect
both the retardation effects and the magnetic field associated with the electric field present in the media. Thus, in both
media, the electric field is given by $\mathbf{E}=-\nabla\phi(\mathbf{r},\omega)$,
where $\phi(\mathbf{r},\omega)$ is the electric scalar potential.
Since the
location of the dipole along the $xy$ plane can be chosen arbitrarily, it is
appropriate to perform a partial Fourier transform from real space to the reciprocal
space of the wavevector $\mathbf{q}=(q_x,q_y)$, keeping however the dependency of $\phi$
on the $z$ coordinate~\cite{Persson-Lang}.

In medium 1, the electric potential obeys Poisson's equation with
a source term that describes the presence of the polarizable nanoparticle, modeled
as an electric dipole of magnitude $\boldsymbol\mu$, which we will later set to be
proportional to the applied field. In medium 2, the electric potential obeys Laplace's
equation. One relates the field in the two media through the boundary condition that determines the discontinuity in the normal component of the electric displacement vector, $\mathbf{D}=\varepsilon_0\,\varepsilon\,\mathbf{E}$, across the interface. Such a boundary condition reads, after Fourier transformation,
\begin {eqnarray}
\label{eqbci}
\nonumber
\varepsilon_1(\omega)\left.\frac{\partial
\phi(z,\mathbf{q},\omega)}{\partial z}\right|_{z=0^+}
\,-\,\varepsilon_2(\omega)\left.\frac{\partial\phi(z,\mathbf{q},
\omega)}{\partial z}\right|_{z=0^-}\\
=-\frac {\delta\rho(\mathbf{q},\omega)}{\varepsilon_0}\,,
\end {eqnarray}
where $\varepsilon_0$ is the vacuum permittivity and $\delta\rho(\mathbf{q},\omega)$ is the surface density of charge
induced in the graphene sheet. It is related to the current density, $\boldsymbol{\jmath}(\mathbf{q},\omega)$, in the graphene sheet
by the continuity equation
\beq
\label{cont}
\omega \,\delta\rho(\mathbf{q},\omega)=\mathbf{q}\cdot\boldsymbol{\jmath}(\mathbf{q},\omega)\,.
\eeq
In order to close the system of equations necessary for the solution of the problem,
we need the equation that relates the surface density of current to the local electric
field present in the graphene sheet. Within the realm of a diffusion-drift model that
describes the graphene sheet, one has
\beq
\label{constj}
\boldsymbol{\jmath}(\mathbf{q},\omega)=-i\sigma(\omega)\mathbf{q}\left(
\phi(z=0,\mathbf{q},\omega)+\frac{1}{e^2}\frac{\partial E_F}{\partial n}
\frac{\delta\rho(\mathbf{q},\omega)}{1-i\omega\tau_F}\right),
\eeq
where the derivative $\frac{\partial E_F}{\partial n}$, with $E_F$ being
the Fermi energy of graphene and $n$ the density of carriers, is computed
at thermal equilibrium and is given, at zero temperature, by $\frac{\partial E_F}{\partial n}=
\rho^{-1}(E_F)$; i.e., this quantity is just the inverse density of states
of graphene at the Fermi level. Finally, $\tau_F$ denotes the quasiparticles'
relaxation time that enters the Drude formula of the conductivity (see below).
This formula also shows that it is possible to neglect the diffusion term with respect
to the drift one in the limit of $\omega \tau_F\gg 1$.

Substituting \eqref{constj} in \eqref{cont} and introducing the diffusion constant of carriers
in graphene through
\beq
D(\omega)=\frac{\sigma(\omega)}{e^2(1-i\omega\tau_F)}\,\frac{\partial E_F}{\partial n}
=\frac{v_F^2\tau_F}{2(1-i\omega\tau_F)^2}\,,
\label{eqDo}
\eeq
where we have used the Drude form for the conductivity of graphene\cite{c:primer}, $\sigma(\omega)=\sigma_0/(1-i\omega\tau_F)$, with $\sigma_0=\frac{1}{2}e^2
v_F^2\tau_F\rho(E_F)$, with $v_F$ being graphene's Fermi's velocity, we obtain
\beq
\label{denpot}
\delta\rho(\mathbf{q},\omega)=-\frac{\sigma(\omega)\,q^2}{-i\omega+D(\omega)\,q^2}\,\phi(z=0,\mathbf{q},\omega)\,,
\eeq
which relates the local density of charge in the graphene sheet with the local value of
the electric potential. Substituting this equation in \eqref{eqbci}, we obtain the relation
\begin {eqnarray}
\label{eqbc2}
\nonumber
\varepsilon_1(\omega)\left.\frac{\partial
\phi(z,\mathbf{q},\omega)}{\partial z}\right|_{z=0^+}
\,-\,\varepsilon_2(\omega)\left.\frac{\partial\phi(z,\mathbf{q},
\omega)}{\partial z}\right|_{z=0^-}\\
=\frac{\sigma(\omega)\,q^2}{\varepsilon_0\left(-i\omega+D(\omega)\,q^2\right)}\,\phi(z=0,\mathbf{q},\omega)\,,
\end {eqnarray}
which is in a form that involves the electric potential alone. One needs to add to (\ref{eqbc2}) the condition of continuity of the potential at the graphene sheet,
\beq
\label{eqbc3}
\phi(z=0^-,\mathbf{q},\omega)=\phi(z=0^+,\mathbf{q},\omega)\,,
\eeq
equivalent to the condition of continuity of the transverse components of the electric field and necessary for \eqref{eqbc2} to be properly defined. The solution of Poisson's
equation in medium 1, with a source term representing the electric dipole, and of
Laplace's equation in medium 2, with both solutions satisfying the boundary conditions \eqref{eqbc2}
and \eqref{eqbc3}, constitutes the mathematical solution of our physical problem, to which
we turn to in the next section.

\section{Solution by method of images in reciprocal space}
\label{secB}
In the absence of the graphene sheet, the problem described above is solvable through
the method of images.\cite {Jackson} In medium 1, the electric potential is given by the superposition
of the potential created by the original dipole and that of an image-dipole of appropriate
strength,
$\boldsymbol{\mu}^\prime =A[2(\boldsymbol{\mu}\cdot\hat{\mathbf{z}})\hat{\mathbf{z}}-\boldsymbol{\mu}]$, located at $-\mathbf{r}_0=(0,0,-h)$, where $\hat{\mathbf{z}}$ is the unit vector in the direction perpendicular to the interface. In medium 2, the
potential is that of a dipole placed at $\mathbf{r}_0$ in medium 1,
but with a strength that is different in
magnitude from that of the original dipole, $\boldsymbol{\mu}^{\prime \prime} =B\boldsymbol{\mu}$. In the mixed real-space/reciprocal-space representation
used above, the solution of the problem in the absence of the graphene sheet is given,
in regions 1 and 2, by
\begin{eqnarray}
\label{eqphi1}
\phi_1(z,\mathbf{q},\omega)=\frac{1}{2\varepsilon_0\varepsilon_1}\,
\left[A\,\left[\,(\boldsymbol{\mu}\cdot\hat{\mathbf{z}})+i\boldsymbol{\mu}\cdot\hat{\mathbf{q}}\,\right]\,e^{-q(z+h)}\,\right.\nonumber \\
\left. +\,\left[\,(\boldsymbol{\mu}\cdot\hat{\mathbf{z}})\,\mbox{sgn}(z-h)-i\boldsymbol{\mu}\cdot\hat{\mathbf{q}}\,\right]\,e^{-q\mid z-h\mid}
\right]\,,\\
\label{eqphi2}
\phi_2(z,\mathbf{q},\omega)=-\frac{B}{2\varepsilon_0 \varepsilon_2}\,
\left[\,(\boldsymbol{\mu}\cdot\hat{\mathbf{z}})+i\boldsymbol{\mu}\cdot\hat{\mathbf{q}}\,\right]\,e^{q(z-h)}\,,
\end{eqnarray}
where $\hat{\mathbf{q}}$
is the unit vector along $\mathbf{q}$.

In order to generalize this solution to the case where the graphene sheet is present at the
interface, all one has to do is to consider the coefficients $A$ and
$B$ as functions of $\mathbf{q}$. Substituting the solutions \eqref{eqphi1} and \eqref{eqphi2} in the boundary conditions \eqref{eqbc2}
and \eqref{eqbc3} yields
\begin{eqnarray}
\label{eqAqw}
A(q,\omega)&=&\frac{\varepsilon_2-\varepsilon_1+f(q,\omega)}{\varepsilon_1 +\varepsilon_2+f(q,\omega)}\:,
\\
\label{eqBqw}
B(q,\omega)&=&\frac{2\,\varepsilon_2}{\varepsilon_1 +\varepsilon_2+f(q,\omega)}\:,
\end{eqnarray}
where
\begin{equation}
\label{eqfqw}
f(q,\omega)=\frac {q\sigma (\omega)}{\varepsilon_0 \left(-i\omega+D(\omega)\,q^2\right )}\,.
\end{equation}
Equations \eqref{eqphi1} and \eqref{eqphi2}, with
$A(q,\omega)$ and $B(q,\omega)$ given, respectively, by \eqref{eqAqw} and \eqref{eqBqw},
constitute the general solution of the considered problem within the realm of the electrostatic approximation.

\section{Renormalized polarizability}
\label{secC}
\subsection{General expressions}
\label{subsec:expressions}

We now consider that a homogeneous electric field $\mathbf{E}^0(\omega)$
is applied to the system. In the absence of the graphene sheet and for $\varepsilon_2=\varepsilon_1$,
the particle would respond to such a field by developing an electric dipole moment $\boldsymbol{\mu}=\varepsilon_1 \alpha_0(\omega)\,\mathbf{E}^0$, where $\alpha_0(\omega)$ is the particle's polarizability,\cite{Novotny-Hecht} which depends on the material
nature and geometry of the particle as well as on $\varepsilon_1$. We shall consider it as a scalar function of frequency. In the situation depicted in Fig. \ref {fig1}, we have
\beq
\label{renp0}
\boldsymbol{\mu}=\varepsilon_1 \alpha_0(\omega)\,\mathbf{E}^l\,,
\eeq
where $\mathbf{E}^l=\mathbf{E}^0-\mathbf{\nabla}\phi_{pol}$ and $\phi_{pol}$ is the potential created by the polarization charges at the interface, i.e. excluding the self-field created by the nano-particle,
which is represented by the first term of Eq.\eqref{eqphi1}.
Expressing the dipole moment (\ref{renp0}) in terms of $\mathbf{E}^0$, we define the renormalized polarisability\cite{Wind1987}
\beq
\label{renp}
\boldsymbol{\mu}=\varepsilon_1 \boldsymbol{\alpha}^*(\omega)\cdot \,\mathbf{E}^0\,,
\eeq
where the quantity $\boldsymbol{\alpha}^*(\omega)$ is a second-rank tensor.
The electric potential $\phi_{pol}$ is given, in real space and in medium 1, by
\begin{eqnarray}
\label{potrs}
\phi_{pol}(\mathbf{r},\omega)&=&\frac{1}{2\varepsilon_0\varepsilon_1}\,\int\,\frac{d^2q}{(2\pi)^2}\,e^{i\mathbf{q}\cdot\boldsymbol{\rho}} \nonumber \\
&&\times A(q,\omega)\,[\,(\boldsymbol{\mu}\cdot\hat{\mathbf{z}})+i\boldsymbol{\mu}\cdot\hat{\mathbf{q}}\,]\,e^{-q(z+h)}\,,
\end{eqnarray}
where $\boldsymbol{\rho}=(x,y)$ and $A(q,\omega)$ is given by \eqref{eqAqw}.

Applying the gradient operator under the integration sign and substituting \eqref{renp0}, we obtain for the local field acting on the particle
\begin{eqnarray}
\label{El2}
\mathbf{E}^l(\omega)&=&\mathbf{E}^0(\omega)
+\frac{\alpha_0(\omega)}{2\varepsilon_0}
\,\int\,\frac{d^2q}{(2\pi)^2}\,q\,e^{-2qh}\,
\,A(q,\omega)\,\nonumber \\
&&\times \left (\,E^l_z\hat{\mathbf{z}}+(\mathbf{E}^l\cdot\hat{\mathbf{q}})\,\hat{\mathbf{q}}\,\right)
\,.
\end{eqnarray}
Since $A(q,\omega)$ only depends on the modulus of $\mathbf{q}$, one can easily perform the angular
integrals in \eqref{El2}, which yields
\beq
\label{El3}
\mathbf{E}^l(\omega)=\mathbf{E}^0(\omega)+\frac{\alpha_0(\omega)}{8\pi \varepsilon_0}\,
a(h,\omega)\,(\,2E_z^l\hat{\mathbf{z}}+E_x^l\hat{\mathbf{x}}+E_y^l\hat{\mathbf{y}}\,)
\,,
\eeq
where $\hat{\mathbf{x}}$ and $\hat{\mathbf{y }}$ are the unit vectors in the directions of the interface and
\beq
\label{El4}
a(h,\omega)=\int_0^\infty\,dq\,q^2 e^{-2qh}A(q,\omega)\:.
\eeq
Using the components of this equation to express $\mathbf{E}^l$ in terms of $\mathbf{E}^0$,
substituting in \eqref{renp0} and comparing with \eqref{renp}, we obtain the following expressions for the principal components
of the tensor $\boldsymbol{\alpha}^*\!$:
\begin{eqnarray}
\label{axy}
\alpha^*_{xx}(\omega)=\alpha^*_{yy}(\omega)=\frac{\alpha_0(\omega)}{1-\frac{\alpha_0(\omega)}{8\pi \varepsilon_0}\,a(h,\omega)}\,,\\
\label{azz}
\alpha^*_{zz}(\omega)=\frac{\alpha_0(\omega)}{1-\frac{\alpha_0(\omega)}{4\pi \varepsilon_0}\,a(h,\omega)}\,.
\end{eqnarray}
At sufficiently high frequencies, we can neglect the diffusion term
[i.e., we can set $D(\omega)=0$] in the expressions
above, reducing $a(h,\omega)$ to the following form:
\begin{eqnarray}
\label{ahw2}
a(h,\omega)&=&\frac{1}{4 h^3}\left(\, 1+\frac{\beta_1}{1+\varepsilon_2/\varepsilon_1}
+\frac{\beta_1^ 2}{1+\varepsilon_2/\varepsilon_1}\right.\nonumber \\
&&\left.+\frac{\beta_1^3}{1+\varepsilon_2/\varepsilon_1}\,
e^{-\beta_1}\,[-\mbox{Ei}(\beta_1)+i\pi]\,\right)\,,
\end{eqnarray}
where $\beta_1=i\frac{2\omega\varepsilon_0\,h(\,\varepsilon_1+\varepsilon_2\,)}{\sigma (\omega)}$
and $\mbox{Ei}(\beta_1)=-P\int_{-\beta_1}^\infty\,dx\,\frac{e^{-x}}{x}$ is the exponential integral function.\cite{Abramowitz}

We would like to point out the connection between the expression obtained for the renormalized polarizability and
the existence of surface plasmon polaritons (SPPs) in graphene.
The integral appearing in the definition of the function $a(h,\omega)$, Eq. (\ref{El4}) has the form\cite{Note1}

\begin {equation}
\label{integral}
I=\int _0 ^\infty \frac{\varepsilon_2-\varepsilon_1+f(q,\omega)}{\varepsilon_1 +\varepsilon_2+f(q,\omega)}\:e^{-2qh}\!q^2dq
\,.
\end{equation}
Neglecting the diffusion term, the poles of the integrand are given by the equation
\begin{equation}
\frac{\varepsilon_2+\varepsilon_1}q = \frac {\sigma (\omega )}{i\omega \varepsilon_0}
\,,
\label{eq:SPP}
\end{equation}
which is the SPP dispersion relation in the electrostatic approximation.~\cite {c:primer} The SPP wavevector for
a given $\omega $, $q_1$, determines the dependence of $a(h,\omega)$ upon the distance between the particle and
the graphene sheet, $h$, since $\beta_1 =2 q_1 h$ in Eq. (\ref{ahw2}).
\subsection{Examples}
\label{subsec:examples}
For a spherical particle of a radius $R$, made of a dispersive material with a dielectric function $\varepsilon_3(\omega)$, we have~\cite{Jackson}
\beq
\label{eq:bare-pol}
\alpha_0(\omega)=4\pi \varepsilon_0 \frac {\varepsilon_3(\omega)-\varepsilon_1}{\varepsilon_3(\omega)+2\varepsilon_1}R^3
\,.
\eeq
This formula can describe a simple dielectric inclusion, a metallic particle,~\cite{Blackman} if $\varepsilon_3(\omega)=\varepsilon _\infty + i\sigma_{3D}(\omega)/(\varepsilon_0 \omega)$ with $\varepsilon _\infty$ denoting the background dielectric permittivity and $\sigma_{3D}$ the optical conductivity of a bulk metal and also, with some modification, a semiconductor QD.~\cite{Bludov2012}
For a particle made of a typical metal, such as gold, with the plasma frequency lying in the UV spectral region, we have $\varepsilon_3(\omega)\gg 1$ and $\alpha_0\approx 4\pi \varepsilon _0 R^3$ in the THz range, i.e., the bare polarizability is nearly real and dispersionless.

Note that for a spherical particle, one obtains, by setting $\sigma(\omega)=0$
in (\ref{eqAqw}) and substituting the result in (\ref {El4}) (i.e., in the absence of graphene), the formula
\begin{equation}
a(h,\omega)=\frac 1{4h^3} \frac {\varepsilon_2-\varepsilon_1}{\varepsilon_2+\varepsilon_1}
\label{eq:a2}
\end{equation}
and the expressions for the components of $\mathbf{\alpha^*}$, (\ref{axy}) and (\ref{azz}), coincide with those obtained in Ref. \onlinecite{Wind1987}.

\begin{figure}
\begin{center}
\includegraphics[width=8.5cm]{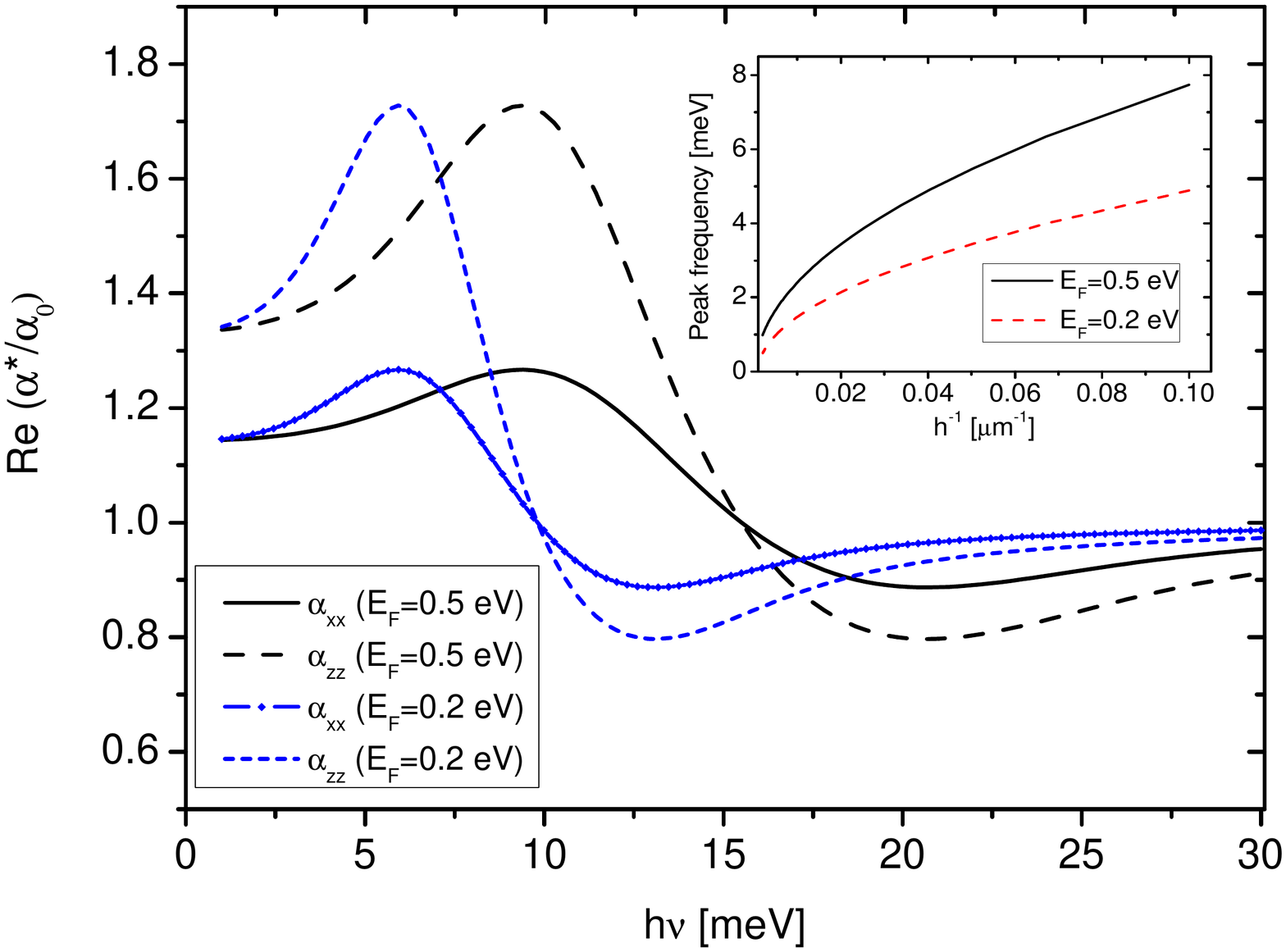}
\end{center}
\caption{(Color online) Frequency dependence of the real part of the renormalized polarizability components (divided by the bare one, $\alpha _{0}$) for a spherical Au particle of radius $R= 10\:\mu$m located at a distance $h= 10\:\mu$m from a suspended graphene sheet ($\varepsilon_2=\varepsilon_1=1$), for two values of the Fermi level as indicated. The inset shows the peak frequency dependence on the inverse of the particle's distance from the graphene sheet. Gold parameters are taken from Ref.~\onlinecite{Torell}.}
\label{fig2}
\end{figure}

Considering the Drude form of the optical conductivity of graphene,
introduced in Eq.\eqref{eqDo} above,
we have that, in the limit of high frequencies, the real part of the conductivity
is small with respect to the imaginary part. Assuming also that $\varepsilon_1$
and $\varepsilon_2$ are real constants, the dispersion relation (\ref{eq:SPP}) yields $\omega \propto \sqrt q$.
The renormalized polarizability of a gold particle of several microns in size is shown in Fig. \ref {fig2}. It shows a Lorentzian-type dispersion induced by the polarization of graphene, larger for the "normal" ($zz$) component.
The position of the peak depends on the distance and the dependence on $h^{-1}$ resembles the SPP dispersion as can be seen in the inset of Fig. \ref {fig2}.
We can say that the renormalized polarizability presents a resonance due to the excitation of SPPs in graphene, with the wavevector $q \sim h^{-1}$.
The other components of $\mathbf{\alpha^*}$ show a similar behavior but the amplitude of the resonance is smaller.
The same conclusions are valid for particles made of a dispersionless dielectric or even for a spherical cavity in one of the dielectrics surrounding the graphene sheet; however, the coupling is weaker in these cases.

\begin{figure}
\begin{center}
\includegraphics[width=8.5cm]{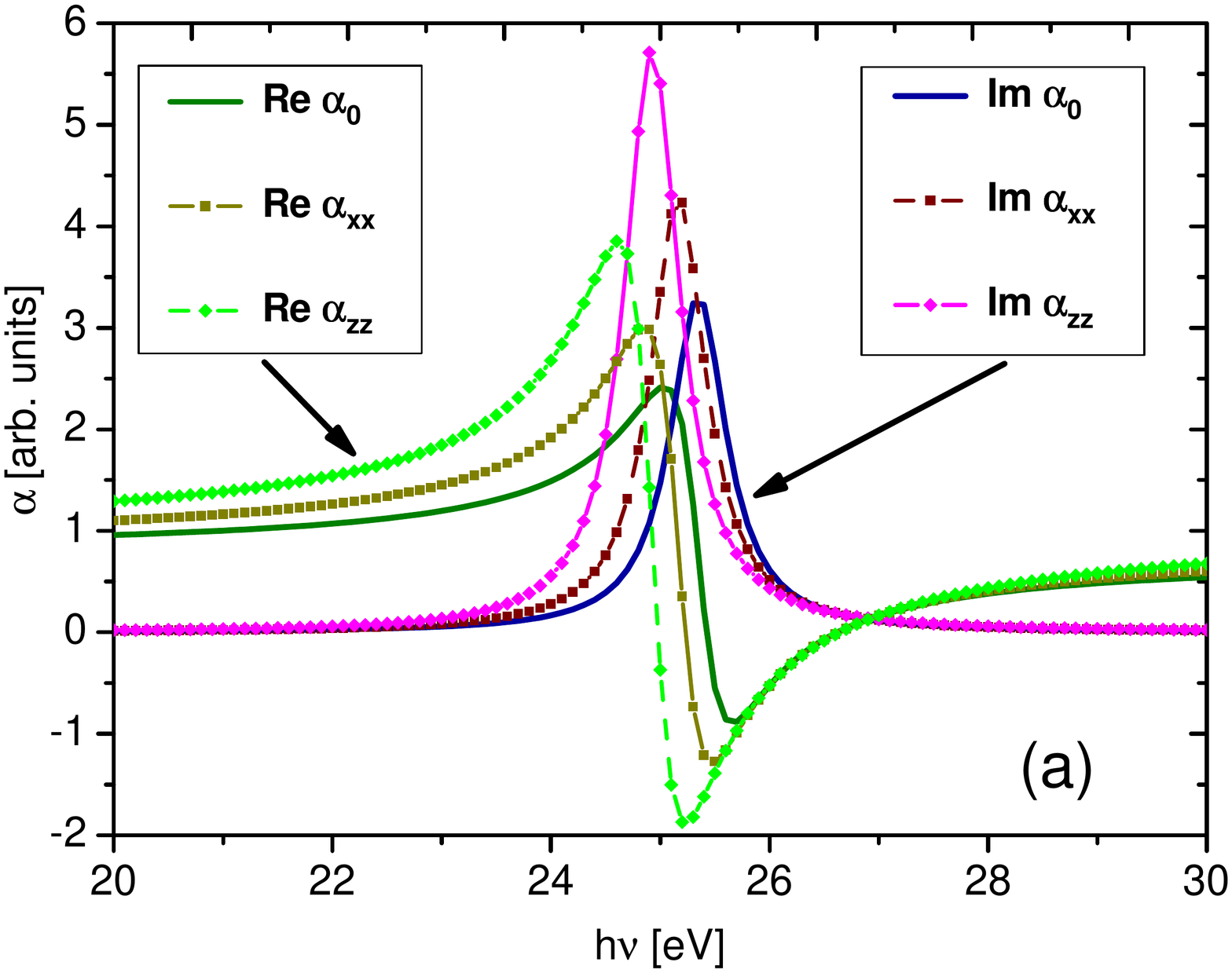}
\includegraphics[width=8.5cm]{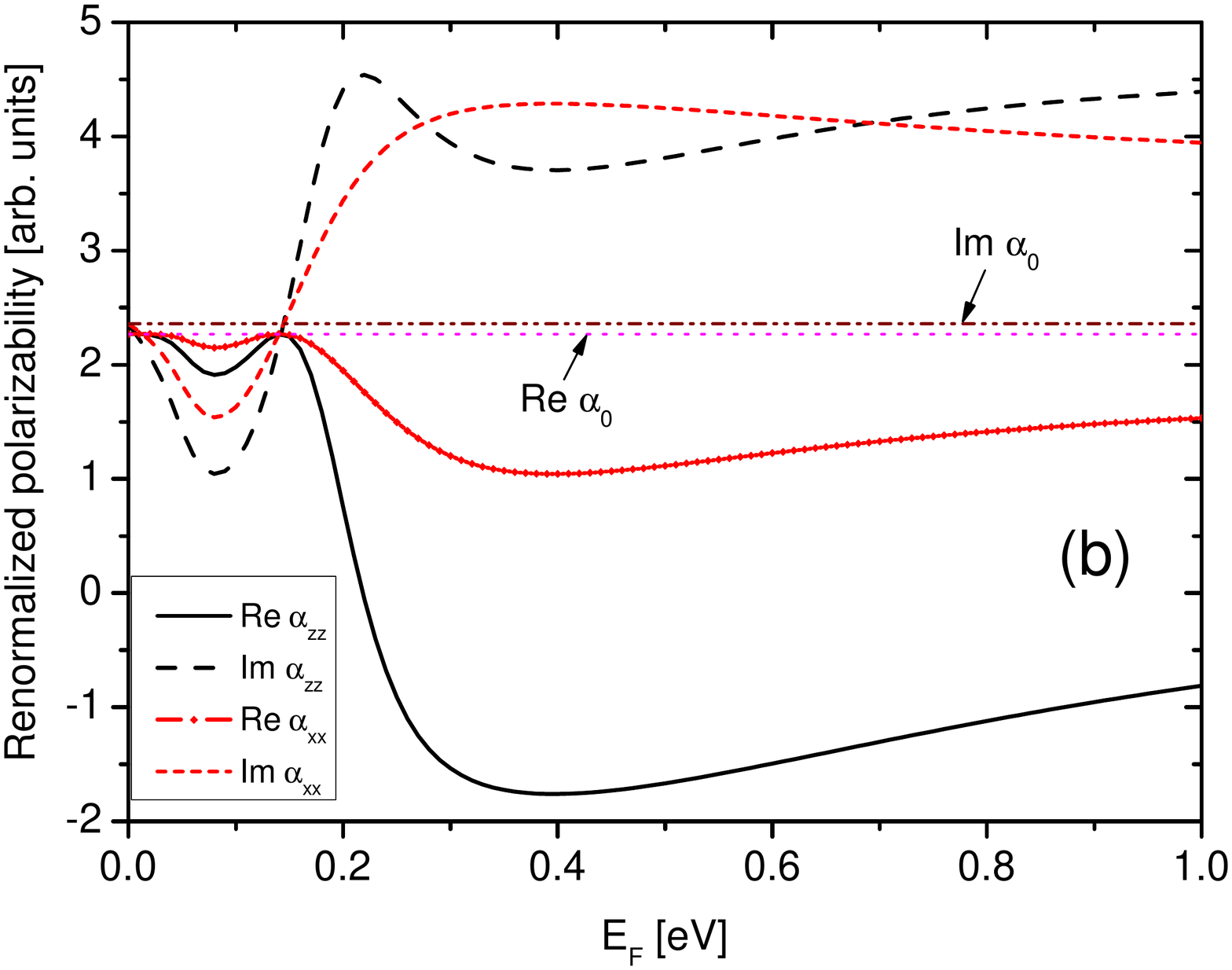}
\end{center}
\caption{(Color online) Dependence of the real and imaginary parts of the renormalized polarizability components on the frequency (a) and the Fermi level (b), calculated for a spherical CdSe particle of radius $R= 1\:\mu$m located at a distance $h= 1.1\:\mu$m from a suspended graphene sheet ($\varepsilon_2=\varepsilon_1=1$). The graphene Fermi level is $E _F=0.4$~eV for (a) and
the field frequency is $\hbar \omega _0=25.15$~meV for (b). Panel (a) also shows the frequency dependence of the bare polarizability, $\alpha _{0}$.
Thin horizontal lines in panel (b) indicate the values of the real and imaginary parts of $\alpha _{0}$ for $\hbar \omega _0=25.15$~meV. Phonon parameters of CdSe were taken from Ref.~\onlinecite{Hamma}. }
\label{fig3}
\end{figure}

An interesting situation arises when the particle's bare polarizability has its own resonance; for instance, if it is made of a polar semiconductor, for example CdSe, with a characteristic reststrahlen band between the transverse ($\omega _{TO}$) and longitudinal ($\omega _{LO}$) optical phonon frequencies, with the dielectric function given by~\cite{Yu-Cardona}
\begin{equation}
\label{DF}
\varepsilon _3 (\omega )=\varepsilon _{\infty}\left (1+ \frac {\omega _{L\!O}^2-\omega
_{TO}^2} {\omega _{TO}^2-\omega ^2-i\omega \Gamma_{TO}}\right ),
\end{equation}
where $\varepsilon _{\infty}=$const and $\Gamma_{TO}$ is the phonon damping. The polarizability of such a particle shows a Lorentzian-type dispersion [see Fig.~\ref{fig3}(a)]. In this case, a double resonance can occur when the SPP frequency (determined by the wavevector $q\sim h^{-1}$) falls within the reststrahlen band (between $\omega _{TO}$ and $\omega _{LO}$) and the denominator in Eq. (\ref{azz}) [or Eq. (\ref{axy})] is small. Although such a resonance is strongly damped because of the large value of Im~$\alpha _0$,
its presence results in a considerable enhancement of the imaginary part of the polarizability, which represents an additional absorption for the particle when located close to the graphene sheet
[see Fig.~\ref{fig3}(a)]. We also note that the peak frequency shifts slightly downwards.

For a fixed frequency, the renormalized polarizability components show a nontrivial dependence upon the Fermi level [see Fig.~\ref{fig3}(b)], with the absorption enhancement taking place above a certain value [$E_F\approx 0.16$~eV in Fig.~\ref{fig3}(b)].
In order to understand this behavior, we recall that the polarizability enhancement factor depends on $E_F$ through Eq. (\ref {ahw2}) and that
\begin{equation}
\beta _1 \approx \frac {2\omega \varepsilon _0(\varepsilon _1 + \varepsilon _2)}{\mbox {Im}\:\sigma} h\,,
\label{eqbetar}
\end{equation}
with $\mbox {Im}\, \sigma \approx 4\alpha _F\, \varepsilon _0\, E_F/(\hbar \omega)$ (where $\alpha _F$ is the fine structure constant), i.e., $\beta _1 \propto (\hbar \omega /E_F)$.
We would recognize in the variation of the real part of the renormalized polarizability seen in Fig. \ref{fig3}(b) the same dispersive behavior seen in  Fig.\ref{fig2}, if we were to represent the functions plotted in this latter figure in terms of $\omega^{-1}$, rather than $\omega$.
The characteristic value of $E_F$ corresponds to the matching of the SPP frequency (at $q \sim h^{-1}$) with the phonon resonance frequency.
In principle, such a pronounced dependence of  $\mbox {Im}\, \alpha ^*$ upon $E_F$ opens the possibility to probe the Fermi level in gated graphene by measuring the resonant absorption of radiation by such particles. Note that such a double resonance should occur whenever $\alpha_0(\omega)$ shows a strong dispersion.
 For instance, for a QD, the real part of the bare polarizability, $\mbox {Re}\,\alpha_0(\omega)$, strongly oscillates in the vicinity of the excitonic transitions and if the dot is made of a narrow gap material (e.g., PbTe) a coupling between a confined QD exciton and surface plasmon waves can take place.~\cite{Bludov2012}

\section{Polarization charge on graphene}
\label{secD}
\begin{figure}
\begin{center}
\includegraphics[width=8.5cm]{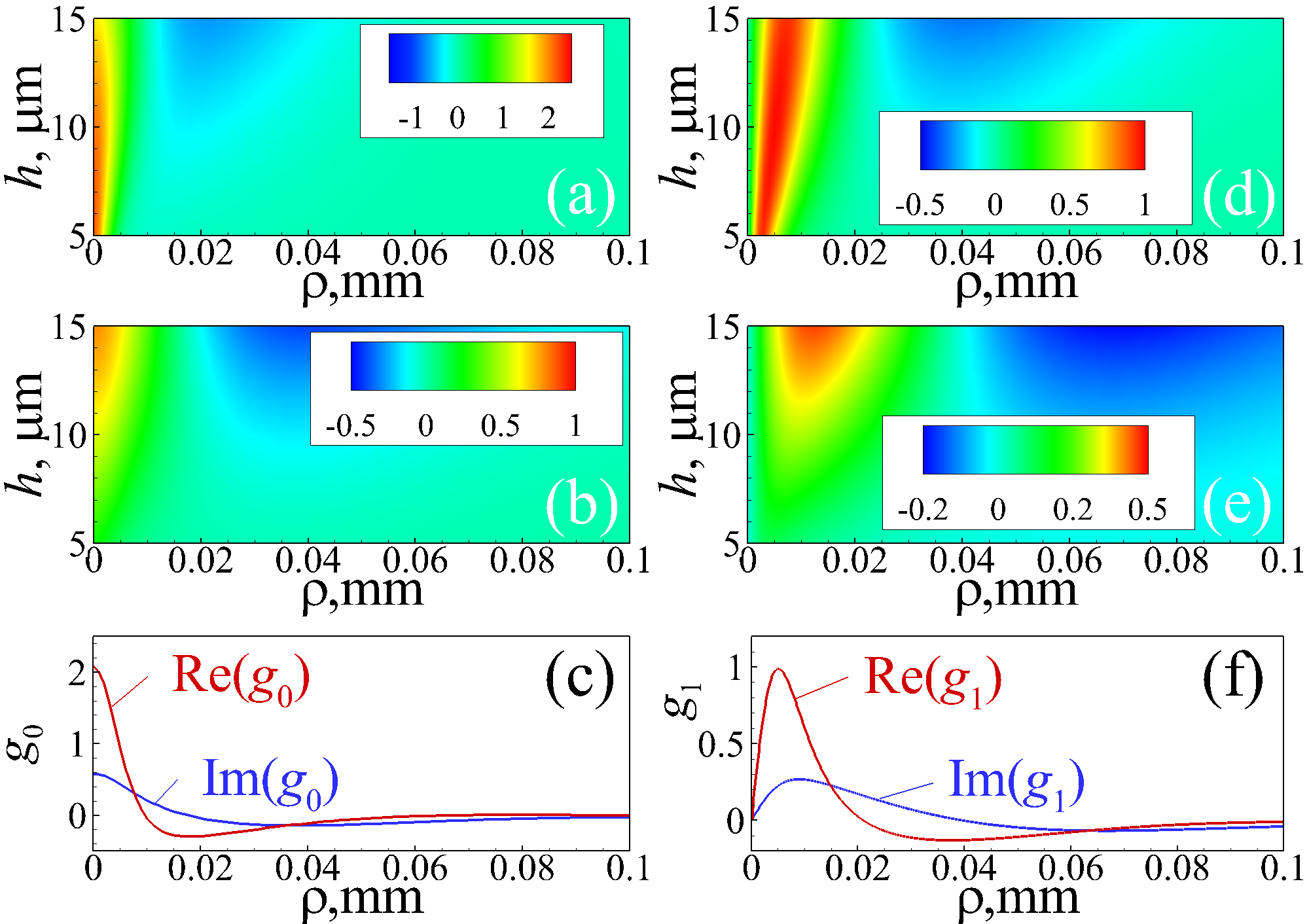}
\end{center}
\caption{(Color online) Dependence of the characteristic functions $g_{0}$ (left column) and $g_{1}$ (right column) that determine the polarization charge density [Eq. (\ref{denrs1})] upon height above the interface and the distance from the dipole projection in the graphene plane. Color code plots show the real [(a) and (d)] and imaginary [(b) and (e)] parts of $g_{0}$ and $g_{1}$ as functions of $h$ and $\rho $.
Plots (c) and (f) are for $h= 10\:\mu$m. The parameters are $\varepsilon_2=\varepsilon_1=1$, $R= 5\:\mu$m, $E_F = 0.5\:$eV, $\tau_F = 10^{-13}\:$s, $\hbar \omega = 10\:$meV.}
\label{fig4}
\end{figure}

Once the renormalized polarizability components are known, one can compute the induced surface charge density $\delta\rho(\boldsymbol{\rho},\omega)$
on the graphene sheet by computing the inverse Fourier transform of
\eqref{denpot}, with $\phi(z=0,\mathbf{q},\omega)$ given by Eq. \eqref{eqphi2}, where $B(q,\omega)$ is defined by \eqref{eqBqw}. This yields,
with $D(\omega)=0$,
\begin{eqnarray}
\label{denrs}
\delta\rho(\boldsymbol{\rho},\omega)
&=&\varepsilon_1\int\,\frac{d^2q}{(2\pi)^2}\,\frac{q^2\,e^{i\mathbf{q}\cdot\boldsymbol{\rho}-qh}}{
q-q_1} \left [\,\alpha^*_{zz}(\omega)E^0_z(\omega)\right . \nonumber \\
&&\left . +i\,\alpha^*_{xx}(\omega)(\hat{\mathbf{q}}\cdot\mathbf{E}^0(\omega))\,\right ]\,,
\end{eqnarray}
where $q_1=\beta _1/(2h)$.
Performing the relevant angular integrals, one obtains
\begin{eqnarray}
\label{denrs1}
\delta\rho(\boldsymbol{\rho},\omega)
&=&\frac{\varepsilon_1}{h^3}\,\left [\,\alpha^*_{zz}(\omega)\,E^0_z(\omega)\,g_0(h,\mathbf{\rho},\omega)\right .\nonumber \\
&&\left .-\,\alpha^*_{xx}(\omega)\,(\mathbf{E}^0_\parallel(\omega)\cdot\hat{\boldsymbol{\rho}})\,g_1(h,\mathbf{\rho},\omega)\,\right ]\,,
\end{eqnarray}
with dimensionless functions $g_{0}$ and $g_{1}$ defined as follows:
\begin{equation}
\label{g12}
g_n(h,\rho,\omega)=\frac{h^3}{2\pi}\int_0^\infty\,dq\,
\frac{q^3 e^{-qh}}{q-q_1}\,J_n(q\rho)\:,
\end{equation}
where $J_n(x)$ is the Bessel function of order $n=0,1$, $\mathbf{E}^0_\parallel(\omega)$ is the applied electric field along the
interface, and $\hat{\boldsymbol{\rho}}$ is the unit vector
along $\boldsymbol{\rho}$. The dependence of the functions $g_{0}$ and $g_{1}$ upon the distance within the graphene plane is shown in Fig.~\ref{fig4}.
Note that the first term in Eq. (\ref{denrs1}) corresponds to an isotropic charge distribution (we can say that it corresponds to an SPP mode with zero angular momentum, $l=0$), while the second one is proportional to the cosine of the angle between $\mathbf{\rho}$ and $\mathbf{E}^0_\parallel$ (we may call it $l=1$ mode).
These oscillations of the charge density are nothing but the surface plasmons with the wavevector $q_1$. As seen from Fig.~\ref{fig4}, the SPP excitation is more efficient if the external field $\mathbf{E}^0$ is normal to the interface, entailing a larger dipole moment and, consequently, a higher surface charge density induced on graphene. The functions $g_n$ decrease rapidly with $h$ (see upper panels in Fig.~\ref{fig4}) and at large distances from the interface this decay is approximately $\sim h^{-3}$.

\section{Optical spectra of a NP--graphene meta-material}
\label{secE}
\begin{figure}
\begin{center}
\includegraphics[width=8.5cm]{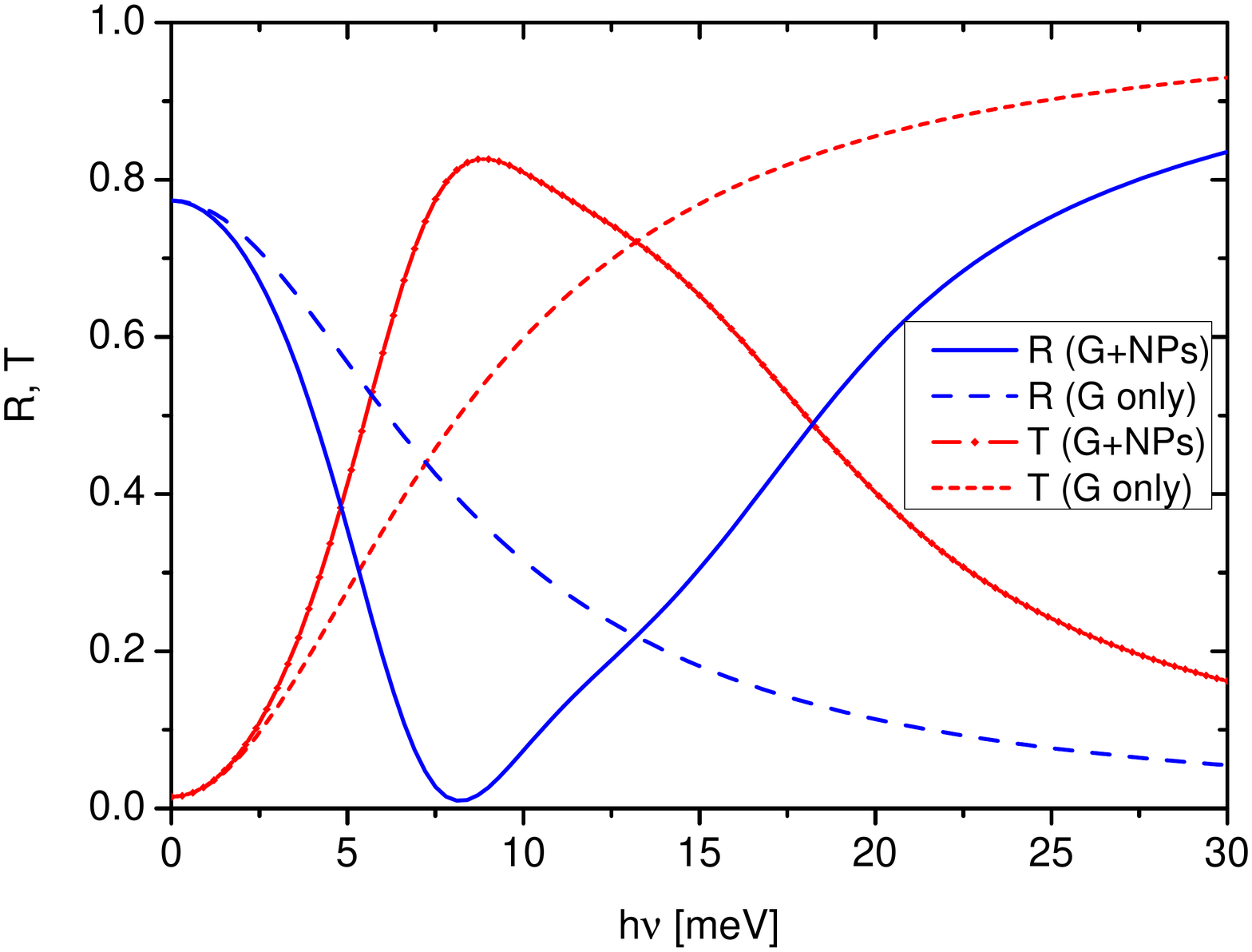}
\includegraphics[width=8.5cm]{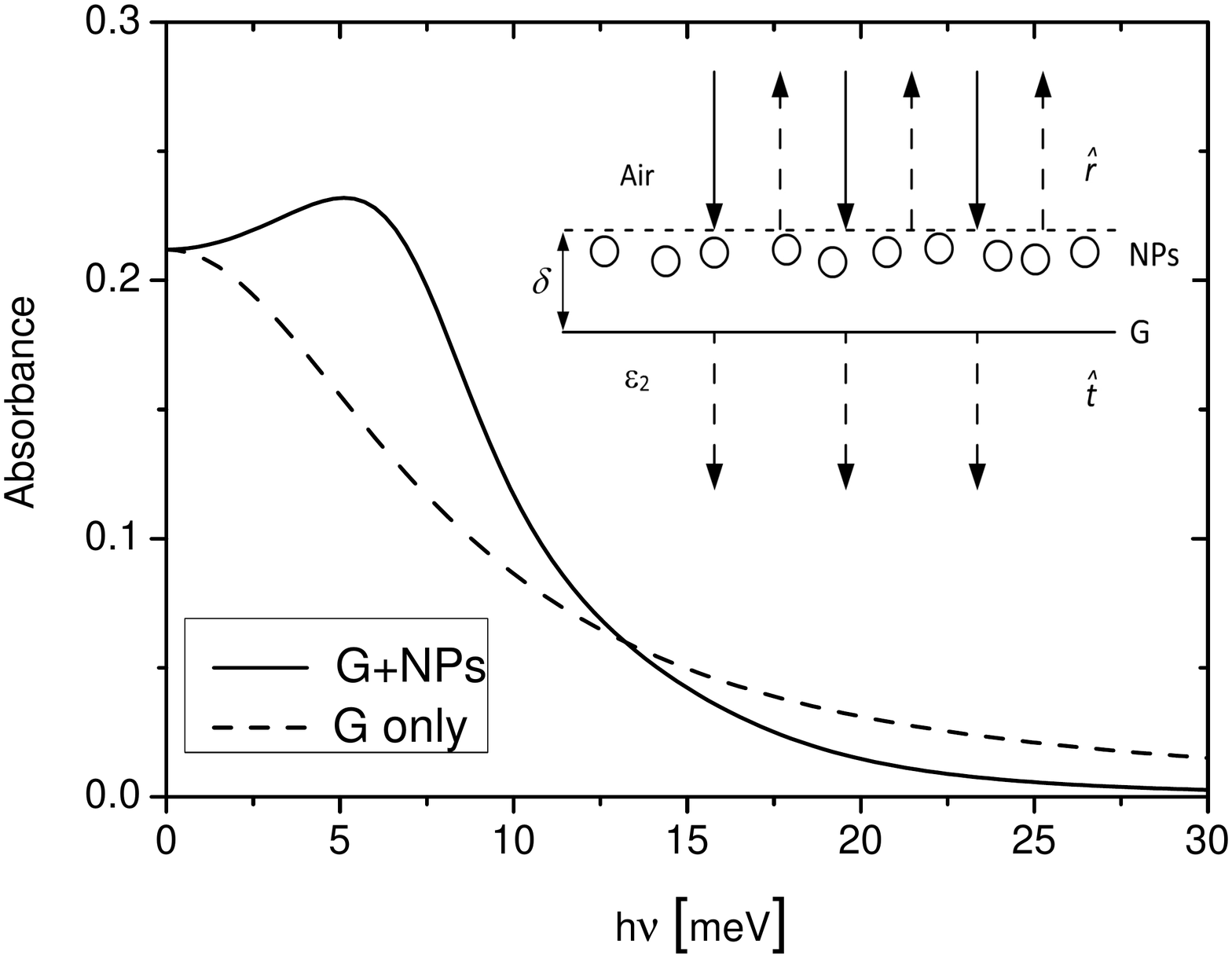}
\end{center}
\caption{(Color online)  Reflectance, transmittance (a) and absorbance (b) spectra of a monolayer of Au NPs randomly deposited on top of a free--standing graphene sheet.
The corresponding spectra of graphene itself are also shown (dashed lines) for comparison. The inset shows the schematics of the considered system. The parameters are the following:
$R= 5\:\mu$m, $h= 5\:\mu$m, $E_F= 0.5\: $eV, $\nu = 8\times 10^{-4}\: \mu$m$^{-2}$.}
\label{fig5}
\end{figure}

Finally, let us consider the situation where polarizable particles are randomly dispersed above a graphene-covered dielectric substrate. For the sake of simplicity, we shall assume that they form a monolayer, i.e., all the particles are located approximately at the same distance ($h$) from the surface. Such monolayers of gold or silver particles can be prepared by colloidal chemistry methods.~\cite{Chumanov}
If the typical distances between them are much larger than $h$, their direct interaction can be neglected and each of the particles can still be described by the renormalized polarizability tensor.
If a plane linear--polarized EM wave impinges the system, at normal incidence (see inset in Fig.~\ref{fig5}), the total dipole moment of the NP layer (per unit area) is simply given by $P_x=\nu\varepsilon_1\alpha^*_{xx}E_x$, where $\nu $ is the number of particles in the monolayer per unit area and $E_x$ is the electric field. In this case, the surface density of the displacement current produced by the time-dependent polarization of the NP layer, $J_x=-i\omega P_x$, can be related to the external field through the effective optical conductivity,
\begin{equation}
\label{sigma_star}
\sigma _{NP} ^*(\omega)=-i\omega \nu \varepsilon_1\alpha^*_{xx}(\omega)\,.
\end{equation}
The polarization current yields a discontinuity of the magnetic component of the EM field ($H_y$), similar to what takes place at a graphene sheet,~\cite{c:primer}
\begin{equation}
\label{bcH}
H_y(z=\delta ^+)-H_y(z=\delta ^-)={\sigma _{NP} ^*} E_x(z=\delta )\:,
\end{equation}
while $E_x$ is continuous across the interface.\cite {Note2} Using these boundary conditions, it is straightforward to obtain the amplitudes of the transmitted and reflected waves (see Appendix \ref{sec:AppB}). In the limit $\omega (R+h)/c \ll 1$ the reflection and transmission coefficients (defined as the ratios of the magnetic field amplitudes) are given by:
\begin{eqnarray}
\label{r_and_t}
\hat r = \frac{1-\sqrt{\varepsilon _2}-\frac {(\sigma +\sigma _{NP} ^*)}{\varepsilon_0 c}}{1+\sqrt{\varepsilon _2}+\frac {(\sigma +\sigma _{NP} ^*)}{\varepsilon_0 c}}\,;
\nonumber \\
\hat t = \frac{2\sqrt{\varepsilon _2}}{1+\sqrt{\varepsilon _2}+\frac {(\sigma +\sigma _{NP} ^*)}{\varepsilon_0 c}}\,.
\end{eqnarray}
The experimentally measured reflectance ($R$) and transmittance of the EM wave are defined as follows:~\cite{Born-Wolf}
\begin{equation}
 \displaystyle R=\left \vert \hat r\right \vert ^2\:,\ \
 \displaystyle T=\frac 1{\sqrt{\varepsilon _2}} \left \vert \hat t\right \vert ^2\:.
\label{RT}
\end{equation}
The absorbance is given by
$
A=1-T-R\:.
$

The quantities $R$, $T$, and $A$ are determined directly by the sum of the optical conductivities of graphene and the NP monolayer, $(\sigma +\sigma _{NP} ^*)$, and the latter takes into account their interaction (the calculated reflectance, transmittance, and absorbance spectra are shown in Fig.~\ref{fig5}). As seen from this figure,
in the vicinity of the SPP resonance (approximately 8 meV in this case) the reflectivity of the structure falls to nearly zero, while the transmittance is increased, compared to the case of pure graphene--covered interface. This effect can be called plasmon-assisted enhanced transmission. Its physical cause (excitation of surface plasmons) is the same of the famous extraordinary optical transmission in metallic films with subwavelength hole arrays~\cite{GarciadeAbajo2007,Ebbesen98}. At the same time, the absorbance is also enhanced in this spectral region and the enhancement factor is nearly 100\% close to the resonance frequency [$A$ increases from $\approx $ 0.13 to $\approx $ 0.23 at $\omega =7$~meV in Fig.~\ref{fig5}(b)] because SPPs in graphene, excited via NPs, are damped.
Note that, as stated above, the bare polarizability of a metallic nanoparticle is (nearly) real. Thus, the rather large imaginary part of its renormalized counterpart is due to the presence of  $a(h,\omega)$ in the denominators of equations \eqref{axy} and \eqref{azz}, and this function only acquires an imaginary part in the presence of graphene, whose coupling to the nanoparticle is responsible for the increased absorbance.
This effect can directly be seen in Eq.~\eqref{r_and_t}, from which one can compute the absorbance of the system: the largest contribution to this quantity comes from the real part of
$\sigma_{NP}^*(\omega)$ [proportional to the imaginary part of $\alpha^*_{xx}(\omega)$], whereas the effect of graphene alone [encoded in $\sigma(\omega)$], is small.
As a result, the absorbance of the whole system is enhanced in the vicinity of $\omega(q_1)$.
As the frequency increases, the reflectivity grows (and $A$ decreases) due to the increasing optical conductivity of the NP layer [see Eq. (\ref{sigma_star})].

\section{Conclusions}
\label{sec_conclusion}

In summary, we derived the electric field created by a polarizable particle located in the vicinity of a graphene-covered interface between two dielectrics. We have shown that the particle's polarizability is renormalized because of its interaction with the polarization charges induced on the interface and, in particular, due to the particle's coupling to surface plasmons supported by graphene.
As a result, the renormalized polarizability is a tensor with two unequal principal components which depend on the frequency even if the bare polarizability is dispersionless.
Since the SPP resonance in graphene is tunable by changing the Fermi level in this material, it is possible to achieve a double resonance with particles possessing resonant bare polarizability, e.g. due to polar optical phonons.
In this case, the absorption of the EM radiation by such particles in the vicinity of the resonance frequency will be enhanced.

In the case of particles whose bare polarizability is frequency--independent in the considered THz range, several effects that are potentially interesting for applications can be achieved, such as
(i) launching of SPPs with metal antennas for nanoscale graphene plasmonic circuits and devices
~\cite{Chen2012,Fei2012,Alonso-Gozalez2014}
[note that a single spherical particle can help launch SPP modes with angular momenta $l=0$ or 1 by choosing an appropriate polarization of the incident wave; see Eq. (\ref {denrs1})]; (ii) scattering and localization of surface plasmons~\cite{Mishchenko};
(iii) enhanced absorption of THz radiation in graphene~\cite{Stauber2014,Echtermeyer};
(iv) enhanced transmission of the EM radiation through doped graphene in a "transparency window" determined by the surface plasmon resonance at a frequency corresponding to the SPP wavevector $q\sim h^{-1}$ ($h$ is the particle's distance from graphene).
These effects can be made broadband if several layers of particles are used for which the resonance frequencies should be somewhat different due to the different distances $h$ of such layers to the graphene sheet.
Finally, we would like to emphasize that, even though we exemplified the predicted effects with micrometer--sized gold spheres, the localized plasmon resonance in these particles is irrelevant and their bare polarizability at the THz frequencies is nearly real and constant, $\alpha _0\approx 4\pi \varepsilon _0 R^3$. In principle, particles made of a dielectric with a sufficiently high permittivity or even spherical holes in a capping dielectric layer could be used instead, even though $\alpha _0$ would be smaller in this case, and one would require higher particle's concentrations to obtain considerable effects. On the theoretical side, it would be necessary to take into consideration particle-particle interactions.~\cite{EPL-Pereira}

\acknowledgments

Financial support from the Portuguese Foundation for Science and Technology (FCT) through Projects PTDC-FIS-113199-2009 and PEst-C/FIS/UI0607/2013 is acknowledged.
We also acknowledge support from the European Commission under Graphene Flagship (Contract No. CNECT-ICT-604391). J.E.S.'s work contract is financed in the framework of the Program of Recruitment of Post Doctoral Researchers for the Portuguese Scientific and Technological System, within the Operational Program Human Potential (POPH) of the
 QREN, participated by the European Social Fund (ESF) and national funds of the Portuguese Ministry of Education and Science (MEC).

\vspace {0.25 cm}
\begin{appendix}
\section{Integral (\ref{integral})}
\label{sec:AppA}
If one does not neglect the diffusion term, the integral (\ref {integral}) is given by
\begin{eqnarray}
I = \frac{A}{4h^3}+\frac{B}{4h^2}-\frac{C}{2h}
+\frac{aq_+-b}{q_+-q_-}e^{-2q_+h}\nonumber \\
\times E_1(-2hq_+)-
\frac{aq_--b}{q_+-q_-}e^{-2q_-h}E_1(-2hq_-)\:,
\end{eqnarray}
where
$$
A=\frac{\varepsilon_2-\varepsilon_1}{\varepsilon_2+\varepsilon_1}\:, \qquad
B=\frac{2\varepsilon_1\sigma}{\varepsilon_0 D(\varepsilon_2+\varepsilon_1)^2}\:,
$$
$$
C=\frac{2\varepsilon_1\sigma^2}{\varepsilon_0^2 D^2(\varepsilon_2+\varepsilon_1)^3}\:,
$$
$$
a=\frac{2\varepsilon_1\sigma ^3}{\varepsilon_0 ^3 D^3(\varepsilon_2+\varepsilon_1)^4}\:,
$$
$$
b=\frac{2i\omega\varepsilon_1\sigma^2}{\varepsilon_0^2 D^3(\varepsilon_2+\varepsilon_1)^3}\:,
$$
and $q_{\pm}$ are the roots of the polynomial
$$
q^2+\frac{\sigma}{\varepsilon_0 D(\varepsilon_2+\varepsilon_1)}q-\frac{i\omega}{D}\:.
$$

\section{Reflection and transmission coefficients}
\label{sec:AppB}

Considering the system schematically shown in the inset in Fig.~\ref{fig5}, we write down the fields in three regions in the following way:

(1) $z\geq \delta =R+h$
\begin{eqnarray}
H_y^{(1)}=e^{-ik(z-\delta)}-\hat r e^{ik(z-\delta)}\:, \nonumber\\
E_x^{(1)}=-e^{-ik(z-\delta)}-\hat r e^{ik(z-\delta)}\:;
\end{eqnarray}

(2) $0\leq z<\delta$
\begin{eqnarray}
H_y^{(2)}=\hat ae^{-ik(z-\delta)}+\hat be^{ik(z-\delta)}\:, \nonumber\\
E_x^{(2)}=-\hat a e^{-ik(z-\delta)}+\hat b e^{ik(z-\delta)}\:;
\end{eqnarray}

(3) $z<0$
\begin{eqnarray}
H_y^{(3)}=\hat te^{-ikz}\:, \nonumber\\
E_x^{(3)}=-\frac {\hat t}{\sqrt {\varepsilon _2}} e^{-ikz}\:.
\end{eqnarray}

By applying the boundary conditions listed in Sec. \ref{secE}, we obtain the following system of equations
for the coefficients $\hat r,\: \hat t,\: \hat a$, and $\hat b$:
$$
1+\hat r=\hat a - \hat b\:,
$$
$$
1-\hat r=\hat a + \hat b +\frac {\sigma _{NP} ^*}{\varepsilon_0 c}\left (1+\hat r \right )\:,
$$
$$
\hat ae^{-ik\delta} - \hat be^{ik\delta}=\frac {\hat t}{\sqrt {\varepsilon _2}}\:,
$$
\begin{equation}
\hat ae^{-ik\delta} + \hat be^{ik\delta}=\hat t + \frac {\sigma }{\varepsilon_0 c}\left (\hat ae^{-ik\delta} - \hat be^{ik\delta} \right )\:.
\label{srt}
\end{equation}
In the limit $k\delta \rightarrow 0$, the system (\ref{srt}) reduces to
\begin{eqnarray}
& & 1+\hat r=\frac {\hat t}{\sqrt {\varepsilon _2}}\:, \nonumber\\
& & 1-\hat r - \hat t=\frac {\hat t}{\sqrt {\varepsilon _2}}\frac {(\sigma _{NP} ^*+\sigma)}{\varepsilon_0 c}\:.
\end{eqnarray}
from which the expressions (\ref{r_and_t}) are obtained.

\end{appendix}


\begin{thebibliography}{55}
\expandafter\ifx\csname natexlab\endcsname\relax\def\natexlab#1{#1}\fi
\expandafter\ifx\csname bibnamefont\endcsname\relax
  \def\bibnamefont#1{#1}\fi
\expandafter\ifx\csname bibfnamefont\endcsname\relax
  \def\bibfnamefont#1{#1}\fi
\expandafter\ifx\csname citenamefont\endcsname\relax
  \def\citenamefont#1{#1}\fi
\expandafter\ifx\csname url\endcsname\relax
  \def\url#1{\texttt{#1}}\fi
\expandafter\ifx\csname urlprefix\endcsname\relax\def\urlprefix{URL }\fi
\providecommand{\bibinfo}[2]{#2}
\providecommand{\eprint}[2][]{\url{#2}}

\bibitem[{\citenamefont{Engheta and Ziolkowski}(2006)}]{Engheta2006}
\bibinfo{editor}{\bibfnamefont{N.}~\bibnamefont{Engheta}} \bibnamefont{and}
  \bibinfo{editor}{\bibfnamefont{R.~W.} \bibnamefont{Ziolkowski}}, eds.,
  \emph{\bibinfo{title}{{Metamaterials - Physics and Engineering
  Explorations}}} (\bibinfo{publisher}{IEEE Press, Piscataway, NJ}, \bibinfo{year}{2006}).

\bibitem[{\citenamefont{Boardman et~al.}(2011)\citenamefont{Boardman,
  Grimalsky, Kivshar, Koshevaya, Lapine, Litchinitser, Malnev, Noginov,
  Rapoport, and Shalaev}}]{Shalaev2011}
\bibinfo{author}{\bibfnamefont{A.}~\bibnamefont{Boardman}},
  \bibinfo{author}{\bibfnamefont{V.}~\bibnamefont{Grimalsky}},
  \bibinfo{author}{\bibfnamefont{Y.}~\bibnamefont{Kivshar}},
  \bibinfo{author}{\bibfnamefont{S.}~\bibnamefont{Koshevaya}},
  \bibinfo{author}{\bibfnamefont{M.}~\bibnamefont{Lapine}},
  \bibinfo{author}{\bibfnamefont{N.}~\bibnamefont{Litchinitser}},
  \bibinfo{author}{\bibfnamefont{V.}~\bibnamefont{Malnev}},
  \bibinfo{author}{\bibfnamefont{M.}~\bibnamefont{Noginov}},
  \bibinfo{author}{\bibfnamefont{Y.}~\bibnamefont{Rapoport}}, \bibnamefont{and}
  \bibinfo{author}{\bibfnamefont{V.}~\bibnamefont{Shalaev}},
  \bibinfo{journal}{Las. Photon. Rev.} \textbf{\bibinfo{volume}{5}},
  \bibinfo{pages}{287} (\bibinfo{year}{2011}).

\bibitem[{\citenamefont{Han and Bozhevolnyi}(2013)}]{Bozhevolnyi2013}
\bibinfo{author}{\bibfnamefont{Z.}~\bibnamefont{Han}} \bibnamefont{and}
  \bibinfo{author}{\bibfnamefont{S.~I.} \bibnamefont{Bozhevolnyi}},
  \bibinfo{journal}{Rep. Prog. Phys.} \textbf{\bibinfo{volume}{76}},
  \bibinfo{pages}{016402} (\bibinfo{year}{2013}).

\bibitem[{\citenamefont{Kravets et~al.}(2008)\citenamefont{Kravets, Schedin,
  and Grigorenko}}]{Kravets2008}
\bibinfo{author}{\bibfnamefont{V.~G.} \bibnamefont{Kravets}},
  \bibinfo{author}{\bibfnamefont{F.}~\bibnamefont{Schedin}}, \bibnamefont{and}
  \bibinfo{author}{\bibfnamefont{A.~N.} \bibnamefont{Grigorenko}},
  \bibinfo{journal}{Phys. Rev. B} \textbf{\bibinfo{volume}{78}},
  \bibinfo{pages}{205405} (\bibinfo{year}{2008}).

\bibitem[{\citenamefont{Ferry et~al.}(2008)\citenamefont{Ferry, Sweadock,
  Pacifici, and Atwater}}]{Ferry}
\bibinfo{author}{\bibfnamefont{V.~E.} \bibnamefont{Ferry}},
  \bibinfo{author}{\bibfnamefont{L.~A.} \bibnamefont{Sweadock}},
  \bibinfo{author}{\bibfnamefont{D.}~\bibnamefont{Pacifici}}, \bibnamefont{and}
  \bibinfo{author}{\bibfnamefont{H.~A.} \bibnamefont{Atwater}},
  \bibinfo{journal}{Nano Lett.} \textbf{\bibinfo{volume}{8}},
  \bibinfo{pages}{4391} (\bibinfo{year}{2008}).

\bibitem[{\citenamefont{Garcia~de Abajo}(2007)}]{GarciadeAbajo2007}
\bibinfo{author}{\bibfnamefont{F.~J.} \bibnamefont{Garcia~de Abajo}},
  \bibinfo{journal}{Rev. Mod. Phys.} \textbf{\bibinfo{volume}{79}},
  \bibinfo{pages}{1267 } (\bibinfo{year}{2007}).

\bibitem[{\citenamefont{Xu et~al.}(2010)\citenamefont{Xu, Wu, Luo, and
  Guo}}]{TingXu2010}
\bibinfo{author}{\bibfnamefont{T.}~\bibnamefont{Xu}},
  \bibinfo{author}{\bibfnamefont{Y.-K.} \bibnamefont{Wu}},
  \bibinfo{author}{\bibfnamefont{X.}~\bibnamefont{Luo}}, \bibnamefont{and}
  \bibinfo{author}{\bibfnamefont{L.~J.} \bibnamefont{Guo}},
  \bibinfo{journal}{Nat. Commun.} \textbf{\bibinfo{volume}{1}},
  \bibinfo{pages}{59} (\bibinfo{year}{2010}).

\bibitem[{\citenamefont{Torrell et~al.}(2010)\citenamefont{Torrell, Cunha,
  Kabir, Cavaleiro, Vasilevskiy, and Vaz}}]{Torell}
\bibinfo{author}{\bibfnamefont{M.}~\bibnamefont{Torrell}},
  \bibinfo{author}{\bibfnamefont{L.}~\bibnamefont{Cunha}},
  \bibinfo{author}{\bibfnamefont{Md.~R.} \bibnamefont{Kabir}},
  \bibinfo{author}{\bibfnamefont{A.}~\bibnamefont{Cavaleiro}},
  \bibinfo{author}{\bibfnamefont{M.~I.} \bibnamefont{Vasilevskiy}},
  \bibnamefont{and} \bibinfo{author}{\bibfnamefont{F.}~\bibnamefont{Vaz}},
  \bibinfo{journal}{Mater. Lett.} \textbf{\bibinfo{volume}{64}},
  \bibinfo{pages}{2014} (\bibinfo{year}{2010}).

\bibitem[{\citenamefont{Kim et~al.}(2010)\citenamefont{Kim, Lee, Yoon, Shin,
  and Shin}}]{Kim}
\bibinfo{author}{\bibfnamefont{K.}~\bibnamefont{Kim}},
  \bibinfo{author}{\bibfnamefont{H.~B.} \bibnamefont{Lee}},
  \bibinfo{author}{\bibfnamefont{J.~K.} \bibnamefont{Yoon}},
  \bibinfo{author}{\bibfnamefont{D.}~\bibnamefont{Shin}}, \bibnamefont{and}
  \bibinfo{author}{\bibfnamefont{K.~S.} \bibnamefont{Shin}},
  \bibinfo{journal}{J. Phys. Chem. C} \textbf{\bibinfo{volume}{114}},
  \bibinfo{pages}{13589} (\bibinfo{year}{2010}).

\bibitem[{\citenamefont{Novotny and Hecht}(2006)}]{Novotny-Hecht}
\bibinfo{author}{\bibfnamefont{L.}~\bibnamefont{Novotny}} \bibnamefont{and}
  \bibinfo{author}{\bibfnamefont{B.}~\bibnamefont{Hecht}},
  \emph{\bibinfo{title}{Principles of Nano-Optics}}
  (\bibinfo{publisher}{Cambridge University Press, Cambridge, UK}, \bibinfo{year}{2006}).

\bibitem[{\citenamefont{Blackman}(2008)}]{Blackman}
\bibinfo{editor}{\bibfnamefont{J.}~\bibnamefont{Blackman}}, ed.,
  \emph{\bibinfo{title}{Metallic Nanoparticles}}
  (\bibinfo{publisher}{Elsevier, New York}, \bibinfo{year}{2008}).

\bibitem[{\citenamefont{{Castro Neto} et~al.}(2009)\citenamefont{{Castro Neto},
  Guinea, Peres, Novoselov, and Geim}}]{rmp}
\bibinfo{author}{\bibfnamefont{A.~H.} \bibnamefont{{Castro Neto}}},
  \bibinfo{author}{\bibfnamefont{F.}~\bibnamefont{Guinea}},
  \bibinfo{author}{\bibfnamefont{N.~M.~R.} \bibnamefont{Peres}},
  \bibinfo{author}{\bibfnamefont{K.~S.} \bibnamefont{Novoselov}},
  \bibnamefont{and} \bibinfo{author}{\bibfnamefont{A.~K.} \bibnamefont{Geim}},
  \bibinfo{journal}{Rev. Mod. Phys.} \textbf{\bibinfo{volume}{81}},
  \bibinfo{pages}{109} (\bibinfo{year}{2009}).

\bibitem[{\citenamefont{Grigorenko et~al.}(2012)\citenamefont{Grigorenko,
  Polini, and Novoselov}}]{novnatphoton}
\bibinfo{author}{\bibfnamefont{A.~N.} \bibnamefont{Grigorenko}},
  \bibinfo{author}{\bibfnamefont{M.}~\bibnamefont{Polini}}, \bibnamefont{and}
  \bibinfo{author}{\bibfnamefont{K.~S.} \bibnamefont{Novoselov}},
  \bibinfo{journal}{Nat. Photon.} \textbf{\bibinfo{volume}{6}},
  \bibinfo{pages}{749} (\bibinfo{year}{2012}).

\bibitem[{\citenamefont{Bludov et~al.}(2013)\citenamefont{Bludov, Ferreira,
  Peres, and Vasilevskiy}}]{c:primer}
\bibinfo{author}{\bibfnamefont{Y.~V.} \bibnamefont{Bludov}},
  \bibinfo{author}{\bibfnamefont{A.}~\bibnamefont{Ferreira}},
  \bibinfo{author}{\bibfnamefont{N.~M.~R.} \bibnamefont{Peres}},
  \bibnamefont{and}
  \bibinfo{author}{\bibfnamefont{M.}~\bibnamefont{Vasilevskiy}},
  \bibinfo{journal}{Int. J. Mod. Phys. B} \textbf{\bibinfo{volume}{27}},
  \bibinfo{pages}{1341001} (\bibinfo{year}{2013}).

\bibitem[{\citenamefont{Luo et~al.}(2013)\citenamefont{Luo, Qiu, Lu, and
  Ni}}]{Luo-r2013}
\bibinfo{author}{\bibfnamefont{X.}~\bibnamefont{Luo}},
  \bibinfo{author}{\bibfnamefont{T.}~\bibnamefont{Qiu}},
  \bibinfo{author}{\bibfnamefont{W.}~\bibnamefont{Lu}}, \bibnamefont{and}
  \bibinfo{author}{\bibfnamefont{Z.}~\bibnamefont{Ni}}, \bibinfo{journal}{Mater.
  Sci. Eng. R} \textbf{\bibinfo{volume}{74}}, \bibinfo{pages}{351}
  (\bibinfo{year}{2013}).

\bibitem[{\citenamefont{Li et~al.}(2008)\citenamefont{Li, Henriksen, Jiang,
  Hao, Martin, Kim, Stormer, and Basov}}]{Li2008}
\bibinfo{author}{\bibfnamefont{Z.~Q.} \bibnamefont{Li}},
  \bibinfo{author}{\bibfnamefont{E.~A.} \bibnamefont{Henriksen}},
  \bibinfo{author}{\bibfnamefont{Z.}~\bibnamefont{Jiang}},
  \bibinfo{author}{\bibfnamefont{Z.}~\bibnamefont{Hao}},
  \bibinfo{author}{\bibfnamefont{M.~C.} \bibnamefont{Martin}},
  \bibinfo{author}{\bibfnamefont{P.}~\bibnamefont{Kim}},
  \bibinfo{author}{\bibfnamefont{H.~L.} \bibnamefont{Stormer}},
  \bibnamefont{and} \bibinfo{author}{\bibfnamefont{D.~N.} \bibnamefont{Basov}},
  \bibinfo{journal}{Nat. Phys.} \textbf{\bibinfo{volume}{4}},
  \bibinfo{pages}{532} (\bibinfo{year}{2008}).

\bibitem[{\citenamefont{Ju et~al.}(2011)\citenamefont{Ju, Geng, Horng, Girit,
  Martin, Hao, Bechtel, Liang, Zettl, Shen et~al.}}]{Ju2011}
\bibinfo{author}{\bibfnamefont{L.}~\bibnamefont{Ju}},
  \bibinfo{author}{\bibfnamefont{B.}~\bibnamefont{Geng}},
  \bibinfo{author}{\bibfnamefont{J.}~\bibnamefont{Horng}},
  \bibinfo{author}{\bibfnamefont{C.}~\bibnamefont{Girit}},
  \bibinfo{author}{\bibfnamefont{M.} \bibnamefont{Martin}},
  \bibinfo{author}{\bibfnamefont{Z.}~\bibnamefont{Hao}},
  \bibinfo{author}{\bibfnamefont{H.~a.} \bibnamefont{Bechtel}},
  \bibinfo{author}{\bibfnamefont{X.}~\bibnamefont{Liang}},
  \bibinfo{author}{\bibfnamefont{A.}~\bibnamefont{Zettl}},
  \bibinfo{author}{\bibfnamefont{Y.~R.} \bibnamefont{Shen}},
  \bibinfo{author}{\bibfnamefont{F.} \bibnamefont{Wang}},
  \bibinfo{journal}{Nat. Nanotechnol.},
  \bibinfo{pages}{630} (\bibinfo{year}{2011}).

\bibitem[{\citenamefont{Nikitin
  et~al.}(2012{\natexlab{a}})\citenamefont{Nikitin, Guinea, Garcia-Vidal, and
  Martin-Moreno}}]{nikitin_ribbon2012}
\bibinfo{author}{\bibfnamefont{A.~Y.} \bibnamefont{Nikitin}},
  \bibinfo{author}{\bibfnamefont{F.}~\bibnamefont{Guinea}},
  \bibinfo{author}{\bibfnamefont{F.~J.} \bibnamefont{Garcia-Vidal}},
  \bibnamefont{and}
  \bibinfo{author}{\bibfnamefont{L.}~\bibnamefont{Martin-Moreno}},
  \bibinfo{journal}{Phys. Rev. B} \textbf{\bibinfo{volume}{85}},
  \bibinfo{pages}{081405} (\bibinfo{year}{2012}{\natexlab{a}}).

\bibitem[{\citenamefont{Yan et~al.}(2012{\natexlab{a}})\citenamefont{Yan, Li,
  Li, Zhu, Avouris, and Xia}}]{Yan_disks2012}
\bibinfo{author}{\bibfnamefont{H.}~\bibnamefont{Yan}},
  \bibinfo{author}{\bibfnamefont{Z.}~\bibnamefont{Li}},
  \bibinfo{author}{\bibfnamefont{X.}~\bibnamefont{Li}},
  \bibinfo{author}{\bibfnamefont{W.}~\bibnamefont{Zhu}},
  \bibinfo{author}{\bibfnamefont{P.}~\bibnamefont{Avouris}}, \bibnamefont{and}
  \bibinfo{author}{\bibfnamefont{F.}~\bibnamefont{Xia}}, \bibinfo{journal}{Nano
  letters} \textbf{\bibinfo{volume}{12}}, \bibinfo{pages}{3766}
  (\bibinfo{year}{2012}{\natexlab{a}}).

\bibitem[{\citenamefont{Thongrattanasiri
  et~al.}(2012)\citenamefont{Thongrattanasiri, Koppens, and {Garc\'{\i}a de
  Abajo}}}]{Thongrattanasiri2012}
\bibinfo{author}{\bibfnamefont{S.}~\bibnamefont{Thongrattanasiri}},
  \bibinfo{author}{\bibfnamefont{F.~H.~L.} \bibnamefont{Koppens}},
  \bibnamefont{and} \bibinfo{author}{\bibfnamefont{F.~J.}
  \bibnamefont{{Garc\'{\i}a de Abajo}}}, \bibinfo{journal}{Phys. Rev. Lett.}
  \textbf{\bibinfo{volume}{108}}, \bibinfo{pages}{047401}
  (\bibinfo{year}{2012}).

\bibitem[{\citenamefont{Feng et~al.}(2014)\citenamefont{Feng, Wang, A.~E, Liu,
  Ajayan, de~Arquer, Nordlander, Zhu, and Halas}}]{Fang2014}
\bibinfo{author}{\bibfnamefont{Z.}~\bibnamefont{Feng}},
  \bibinfo{author}{\bibfnamefont{Y.}~\bibnamefont{Wang}},
  \bibinfo{author}{\bibfnamefont{S.}~\bibnamefont{A.~E}},
  \bibinfo{author}{\bibfnamefont{Z.}~\bibnamefont{Liu}},
  \bibinfo{author}{\bibfnamefont{P.~M.} \bibnamefont{Ajayan}},
  \bibinfo{author}{\bibfnamefont{F.~P.~G.} \bibnamefont{de~Arquer}},
  \bibinfo{author}{\bibfnamefont{P.}~\bibnamefont{Nordlander}},
  \bibinfo{author}{\bibfnamefont{X.}~\bibnamefont{Zhu}}, \bibnamefont{and}
  \bibinfo{author}{\bibfnamefont{N.~J.} \bibnamefont{Halas}},
  \bibinfo{journal}{Nano Lett.} \textbf{\bibinfo{volume}{14}},
  \bibinfo{pages}{299 } (\bibinfo{year}{2014}).

\bibitem[{\citenamefont{Berman et~al.}(2010)\citenamefont{Berman, Boyko,
  Kezerashvili, Kolesnikov, and Lozovik}}]{Berman2010}
\bibinfo{author}{\bibfnamefont{O.~L.} \bibnamefont{Berman}},
  \bibinfo{author}{\bibfnamefont{V.~S.} \bibnamefont{Boyko}},
  \bibinfo{author}{\bibfnamefont{R.~Y.} \bibnamefont{Kezerashvili}},
  \bibinfo{author}{\bibfnamefont{A.~A.} \bibnamefont{Kolesnikov}},
  \bibnamefont{and} \bibinfo{author}{\bibfnamefont{Y.~E.}
  \bibnamefont{Lozovik}}, \bibinfo{journal}{Physics Letters A}
  \textbf{\bibinfo{volume}{374}}, \bibinfo{pages}{4784} (\bibinfo{year}{2010}).

\bibitem[{\citenamefont{Yan et~al.}(2012{\natexlab{b}})\citenamefont{Yan, Li,
  Chandra, Tulevski, Wu, Freitag, Zhu, Avouris, and Xia}}]{Yan2012}
\bibinfo{author}{\bibfnamefont{H.}~\bibnamefont{Yan}},
  \bibinfo{author}{\bibfnamefont{X.}~\bibnamefont{Li}},
  \bibinfo{author}{\bibfnamefont{B.}~\bibnamefont{Chandra}},
  \bibinfo{author}{\bibfnamefont{G.}~\bibnamefont{Tulevski}},
  \bibinfo{author}{\bibfnamefont{Y.}~\bibnamefont{Wu}},
  \bibinfo{author}{\bibfnamefont{M.}~\bibnamefont{Freitag}},
  \bibinfo{author}{\bibfnamefont{W.}~\bibnamefont{Zhu}},
  \bibinfo{author}{\bibfnamefont{P.}~\bibnamefont{Avouris}}, \bibnamefont{and}
  \bibinfo{author}{\bibfnamefont{F.}~\bibnamefont{Xia}},
  \bibinfo{journal}{Nat. Nanotechnol.} \textbf{\bibinfo{volume}{7}},
  \bibinfo{pages}{330} (\bibinfo{year}{2012}{\natexlab{b}}).

\bibitem[{\citenamefont{Nikitin
  et~al.}(2012{\natexlab{b}})\citenamefont{Nikitin, Guinea, and
  Martin-Moreno}}]{nikitin2012}
\bibinfo{author}{\bibfnamefont{A.~Y.} \bibnamefont{Nikitin}},
  \bibinfo{author}{\bibfnamefont{F.}~\bibnamefont{Guinea}}, \bibnamefont{and}
  \bibinfo{author}{\bibfnamefont{L.}~\bibnamefont{Martin-Moreno}},
  \bibinfo{journal}{Appl. Phys. Lett.} \textbf{\bibinfo{volume}{101}},
  \bibinfo{pages}{151119} (\bibinfo{year}{2012}{\natexlab{b}}).

\bibitem[{\citenamefont{Gaudreau et~al.}(2013)\citenamefont{Gaudreau,
  Tielrooij, Prawiroatmodjo, Osmond, de~Abajo, and Koppens}}]{Gaudreau2013}
\bibinfo{author}{\bibfnamefont{L.}~\bibnamefont{Gaudreau}},
  \bibinfo{author}{\bibfnamefont{K.~J.} \bibnamefont{Tielrooij}},
  \bibinfo{author}{\bibfnamefont{C.~E. D.~K.} \bibnamefont{Prawiroatmodjo}},
  \bibinfo{author}{\bibfnamefont{J.}~\bibnamefont{Osmond}},
  \bibinfo{author}{\bibfnamefont{F.~J.~G.} \bibnamefont{de~Abajo}},
  \bibnamefont{and} \bibinfo{author}{\bibfnamefont{F.~H.~L.}
  \bibnamefont{Koppens}}, \bibinfo{journal}{Nano Lett.}
  \textbf{\bibinfo{volume}{13}}, \bibinfo{pages}{2030} (\bibinfo{year}{2013}).

\bibitem[{\citenamefont{Biehs and Agarwal}(2013)}]{Agarwal2013}
\bibinfo{author}{\bibfnamefont{S.-A.} \bibnamefont{Biehs}} \bibnamefont{and}
  \bibinfo{author}{\bibfnamefont{G.~S.} \bibnamefont{Agarwal}},
  \bibinfo{journal}{Appl. Phys. Lett.} \textbf{\bibinfo{volume}{103}},
  \bibinfo{pages}{243112} (\bibinfo{year}{2013}).

\bibitem[{\citenamefont{Velizhanin and Efimov}(2011)}]{Velizhanin2011}
\bibinfo{author}{\bibfnamefont{K.~A.} \bibnamefont{Velizhanin}}
  \bibnamefont{and} \bibinfo{author}{\bibfnamefont{A.}~\bibnamefont{Efimov}},
  \bibinfo{journal}{Phys. Rev. B} \textbf{\bibinfo{volume}{84}},
  \bibinfo{pages}{085401} (\bibinfo{year}{2011}).

\bibitem[{\citenamefont{Chen et~al.}(2010)\citenamefont{Chen, Berciaud,
  Nuckolls, Heinz, and Brus}}]{Chen2010}
\bibinfo{author}{\bibfnamefont{Z.}~\bibnamefont{Chen}},
  \bibinfo{author}{\bibfnamefont{S.}~\bibnamefont{Berciaud}},
  \bibinfo{author}{\bibfnamefont{C.}~\bibnamefont{Nuckolls}},
  \bibinfo{author}{\bibfnamefont{T.~F.} \bibnamefont{Heinz}}, \bibnamefont{and}
  \bibinfo{author}{\bibfnamefont{L.~E.} \bibnamefont{Brus}},
  \bibinfo{journal}{ACS Nano} \textbf{\bibinfo{volume}{4}},
  \bibinfo{pages}{2964} (\bibinfo{year}{2010}).

\bibitem[{\citenamefont{Konstantatos et~al.}(2012)\citenamefont{Konstantatos,
  Badioli, Osmond, Gaudreau, de~Arquer, Gatti, and Koppens}}]{Konstantatos2012}
\bibinfo{author}{\bibfnamefont{G.}~\bibnamefont{Konstantatos}},
  \bibinfo{author}{\bibfnamefont{M.}~\bibnamefont{Badioli}},
  \bibinfo{author}{\bibfnamefont{J.}~\bibnamefont{Osmond}},
  \bibinfo{author}{\bibfnamefont{L.}~\bibnamefont{Gaudreau}},
  \bibinfo{author}{\bibfnamefont{F.~P.~G.} \bibnamefont{de~Arquer}},
  \bibinfo{author}{\bibfnamefont{F.}~\bibnamefont{Gatti}}, \bibnamefont{and}
  \bibinfo{author}{\bibfnamefont{F.~H.~L.} \bibnamefont{Koppens}},
  \bibinfo{journal}{Nat. Nanotechnol.} \textbf{\bibinfo{volume}{7}},
  \bibinfo{pages}{363} (\bibinfo{year}{2012}).

\bibitem[{\citenamefont{Huidobro et~al.}(2012)\citenamefont{Huidobro, Nikitin,
  Gonz\'alez-Ballestero, Martin-Moreno, and Garc\'ia-Vidal}}]{Huidobro2012}
\bibinfo{author}{\bibfnamefont{P.~A.} \bibnamefont{Huidobro}},
  \bibinfo{author}{\bibfnamefont{A.~Y.} \bibnamefont{Nikitin}},
  \bibinfo{author}{\bibfnamefont{C.}~\bibnamefont{Gonz\'alez-Ballestero}},
  \bibinfo{author}{\bibfnamefont{L.}~\bibnamefont{Martin-Moreno}},
  \bibnamefont{and} \bibinfo{author}{\bibfnamefont{F.~J.}
  \bibnamefont{Garc\'ia-Vidal}}, \bibinfo{journal}{Phys. Rev. B}
  \textbf{\bibinfo{volume}{85}}, \bibinfo{pages}{155438}
  (\bibinfo{year}{2012}).

\bibitem[{\citenamefont{Stauber et~al.}(2014)\citenamefont{Stauber,
  G\'omez-Santos, and de~Abajo}}]{Stauber2014}
\bibinfo{author}{\bibfnamefont{T.}~\bibnamefont{Stauber}},
  \bibinfo{author}{\bibfnamefont{G.}~\bibnamefont{G\'omez-Santos}},
  \bibnamefont{and} \bibinfo{author}{\bibfnamefont{F.~J.~G.}
  \bibnamefont{de~Abajo}}, \bibinfo{journal}{Phys. Rev. Lett.}
  \textbf{\bibinfo{volume}{112}}, \bibinfo{pages}{077401}
  (\bibinfo{year}{2014}).

\bibitem[{\citenamefont{Chen et~al.}(2012)\citenamefont{Chen, Badioli,
  Alonso-Gonz\'alez, Thongrattanasiri, Huth, Osmond, Spasenovi\'c, Centeno,
  Pesquera, Godignon et~al.}}]{Chen2012}
\bibinfo{author}{\bibfnamefont{J.}~\bibnamefont{Chen}},
  \bibinfo{author}{\bibfnamefont{M.}~\bibnamefont{Badioli}},
  \bibinfo{author}{\bibfnamefont{P.}~\bibnamefont{Alonso-Gonz\'alez}},
  \bibinfo{author}{\bibfnamefont{S.}~\bibnamefont{Thongrattanasiri}},
  \bibinfo{author}{\bibfnamefont{F.}~\bibnamefont{Huth}},
  \bibinfo{author}{\bibfnamefont{J.}~\bibnamefont{Osmond}},
  \bibinfo{author}{\bibfnamefont{M.}~\bibnamefont{Spasenovi\'c}},
  \bibinfo{author}{\bibfnamefont{A.}~\bibnamefont{Centeno}},
  \bibinfo{author}{\bibfnamefont{A.}~\bibnamefont{Pesquera}},
  \bibinfo{author}{\bibfnamefont{P.}~\bibnamefont{Godignon}},
  \bibnamefont{et~al.}, \bibinfo{journal}{Nature (London)}
  \textbf{\bibinfo{volume}{487}}, \bibinfo{pages}{77} (\bibinfo{year}{2012}).

\bibitem[{\citenamefont{Fei et~al.}(2012)\citenamefont{Fei, Rodin, Andreev,
  Bao, McLeod, Wagner, Zhang, Zhao, Thiemens, Dominguez et~al.}}]{Fei2012}
\bibinfo{author}{\bibfnamefont{Z.}~\bibnamefont{Fei}},
  \bibinfo{author}{\bibfnamefont{A.~S.} \bibnamefont{Rodin}},
  \bibinfo{author}{\bibfnamefont{G.~O.} \bibnamefont{Andreev}},
  \bibinfo{author}{\bibfnamefont{W.}~\bibnamefont{Bao}},
  \bibinfo{author}{\bibfnamefont{A.~S.} \bibnamefont{McLeod}},
  \bibinfo{author}{\bibfnamefont{M.}~\bibnamefont{Wagner}},
  \bibinfo{author}{\bibfnamefont{L.~M.} \bibnamefont{Zhang}},
  \bibinfo{author}{\bibfnamefont{Z.}~\bibnamefont{Zhao}},
  \bibinfo{author}{\bibfnamefont{M.}~\bibnamefont{Thiemens}},
  \bibinfo{author}{\bibfnamefont{G.}~\bibnamefont{Dominguez}},
  \bibnamefont{et~al.}, \bibinfo{journal}{Nature (London)}
  \textbf{\bibinfo{volume}{487}}, \bibinfo{pages}{82} (\bibinfo{year}{2012}).

\bibitem[{\citenamefont{Akimov et~al.}(2007)\citenamefont{Akimov, Mukherjee,
  Yu, Chang, Zibrov, Hemmer, Park, and Lukin}}]{Akimov2007}
\bibinfo{author}{\bibfnamefont{A.~V.} \bibnamefont{Akimov}},
  \bibinfo{author}{\bibfnamefont{A.}~\bibnamefont{Mukherjee}},
  \bibinfo{author}{\bibfnamefont{C.~L.} \bibnamefont{Yu}},
  \bibinfo{author}{\bibfnamefont{D.~E.} \bibnamefont{Chang}},
  \bibinfo{author}{\bibfnamefont{A.~S.} \bibnamefont{Zibrov}},
  \bibinfo{author}{\bibfnamefont{P.~R.} \bibnamefont{Hemmer}},
  \bibinfo{author}{\bibfnamefont{H.}~\bibnamefont{Park}}, \bibnamefont{and}
  \bibinfo{author}{\bibfnamefont{M.~D.} \bibnamefont{Lukin}},
  \bibinfo{journal}{Nature (London)} \textbf{\bibinfo{volume}{450}},
  \bibinfo{pages}{402} (\bibinfo{year}{2007}).

\bibitem[{\citenamefont{Ray and Lakowitz}(2013)}]{Ray2013}
\bibinfo{author}{\bibfnamefont{K.}~\bibnamefont{Ray}} \bibnamefont{and}
  \bibinfo{author}{\bibfnamefont{J.~R.} \bibnamefont{Lakowitz}},
  \bibinfo{journal}{J. Phys. Chem. C} \textbf{\bibinfo{volume}{117}},
  \bibinfo{pages}{15790} (\bibinfo{year}{2013}).

\bibitem[{\citenamefont{Lunz et~al.}(2012)\citenamefont{Lunz, Zhang, Gerard,
  Gunko, Lesnyak, Gaponik, Susha, Rogach, and Bradley}}]{Lunz2012}
\bibinfo{author}{\bibfnamefont{M.}~\bibnamefont{Lunz}},
  \bibinfo{author}{\bibfnamefont{X.}~\bibnamefont{Zhang}},
  \bibinfo{author}{\bibfnamefont{V.~A.} \bibnamefont{Gerard}},
  \bibinfo{author}{\bibfnamefont{Y.~K.} \bibnamefont{Gunko}},
  \bibinfo{author}{\bibfnamefont{V.}~\bibnamefont{Lesnyak}},
  \bibinfo{author}{\bibfnamefont{N.}~\bibnamefont{Gaponik}},
  \bibinfo{author}{\bibfnamefont{A.~S.} \bibnamefont{Susha}},
  \bibinfo{author}{\bibfnamefont{A.~L.} \bibnamefont{Rogach}},
  \bibnamefont{and} \bibinfo{author}{\bibfnamefont{A.~L.}
  \bibnamefont{Bradley}}, \bibinfo{journal}{J. Phys. Chem. C}
  \textbf{\bibinfo{volume}{116}}, \bibinfo{pages}{26529 }
  (\bibinfo{year}{2012}).

\bibitem[{\citenamefont{Schreiber et~al.}(2014)\citenamefont{Schreiber, Do,
  Roller, Zhang, Schüller, Nickels, Feldmann, and Liedl}}]{Schreiber2014}
\bibinfo{author}{\bibfnamefont{R.}~\bibnamefont{Schreiber}},
  \bibinfo{author}{\bibfnamefont{J.}~\bibnamefont{Do}},
  \bibinfo{author}{\bibfnamefont{E.-M.} \bibnamefont{Roller}},
  \bibinfo{author}{\bibfnamefont{T.}~\bibnamefont{Zhang}},
  \bibinfo{author}{\bibfnamefont{V.~J.} \bibnamefont{Sch\"uller}},
  \bibinfo{author}{\bibfnamefont{P.~C.} \bibnamefont{Nickels}},
  \bibinfo{author}{\bibfnamefont{J.}~\bibnamefont{Feldmann}}, \bibnamefont{and}
  \bibinfo{author}{\bibfnamefont{T.}~\bibnamefont{Liedl}},
  \bibinfo{journal}{Nat. Nanotechnol.} \textbf{\bibinfo{volume}{9}},
  \bibinfo{pages}{74} (\bibinfo{year}{2014}).

\bibitem[{\citenamefont{Gomez et~al.}(2010)\citenamefont{Gomez, Vernon,
  Mulvaney, and Davis}}]{Gomez2010}
\bibinfo{author}{\bibfnamefont{D.~E.} \bibnamefont{Gomez}},
  \bibinfo{author}{\bibfnamefont{R.~C.} \bibnamefont{Vernon}},
  \bibinfo{author}{\bibfnamefont{P.}~\bibnamefont{Mulvaney}}, \bibnamefont{and}
  \bibinfo{author}{\bibfnamefont{T.~J.} \bibnamefont{Davis}},
  \bibinfo{journal}{Nano Lett.} \textbf{\bibinfo{volume}{10}},
  \bibinfo{pages}{274} (\bibinfo{year}{2010}).

\bibitem[{\citenamefont{Bludov and Vasilevskiy}(2012)}]{Bludov2012}
\bibinfo{author}{\bibfnamefont{Y.~V.} \bibnamefont{Bludov}} \bibnamefont{and}
  \bibinfo{author}{\bibfnamefont{M.~I.} \bibnamefont{Vasilevskiy}},
  \bibinfo{journal}{J. Phys. Chem. C} \textbf{\bibinfo{volume}{116}},
  \bibinfo{pages}{13738} (\bibinfo{year}{2012}).

\bibitem[{\citenamefont{Mikhailov and Ziegler}(2007)}]{Mikhailov2007}
\bibinfo{author}{\bibfnamefont{S.~A.} \bibnamefont{Mikhailov}}
  \bibnamefont{and} \bibinfo{author}{\bibfnamefont{K.}~\bibnamefont{Ziegler}},
  \bibinfo{journal}{Phys. Rev. Lett.} \textbf{\bibinfo{volume}{99}},
  \bibinfo{pages}{016803} (\bibinfo{year}{2007}).

\bibitem[{\citenamefont{Persson and Lang}(1982)}]{Persson-Lang}
\bibinfo{author}{\bibfnamefont{B.~N.~J.} \bibnamefont{Persson}}
  \bibnamefont{and} \bibinfo{author}{\bibfnamefont{N.~D.} \bibnamefont{Lang}},
  \bibinfo{journal}{Phys. Rev. B} \textbf{\bibinfo{volume}{26}},
  \bibinfo{pages}{5409} (\bibinfo{year}{1982}).

\bibitem[{\citenamefont{Jackson}(1998)}]{Jackson}
\bibinfo{author}{\bibfnamefont{J.~D.} \bibnamefont{Jackson}},
  \emph{\bibinfo{title}{Classical Electrodynamics}} (\bibinfo{publisher}{J.
  Wiley, New York}, \bibinfo{year}{1998}).

\bibitem[{\citenamefont{Wind et~al.}(1987)\citenamefont{Wind, Vlieger, and
  Bedeaux}}]{Wind1987}
\bibinfo{author}{\bibfnamefont{M.~M.} \bibnamefont{Wind}},
  \bibinfo{author}{\bibfnamefont{J.}~\bibnamefont{Vlieger}}, \bibnamefont{and}
  \bibinfo{author}{\bibfnamefont{D.}~\bibnamefont{Bedeaux}},
  \bibinfo{journal}{Physica A} \textbf{\bibinfo{volume}{141}},
  \bibinfo{pages}{33 } (\bibinfo{year}{1987}).

\bibitem[{\citenamefont{Abramowitz and Stegun}(1972)}]{Abramowitz}
\bibinfo{editor}{\bibfnamefont{M.}~\bibnamefont{Abramowitz}} \bibnamefont{and}
  \bibinfo{editor}{\bibfnamefont{I.~A.} \bibnamefont{Stegun}}, eds.,
  \emph{\bibinfo{title}{Handbook of Mathematical Functions}}
  (\bibinfo{publisher}{Dover, New York}, \bibinfo{year}{1972}).

\bibitem[{Not({\natexlab{a}})}]{Note1}
\bibinfo{note}{The expression for this integral without neglecting the
  diffusion term is given in Appendix \ref{sec:AppA}.}

\bibitem[{\citenamefont{Yu and Cardona}(1996)}]{Yu-Cardona}
\bibinfo{author}{\bibfnamefont{P.~Y.} \bibnamefont{Yu}} \bibnamefont{and}
  \bibinfo{author}{\bibfnamefont{M.}~\bibnamefont{Cardona}},
  \emph{\bibinfo{title}{Fundamentals of Semiconductors}}
  (\bibinfo{publisher}{Springer, Berlin}, \bibinfo{year}{1996}).

\bibitem[{\citenamefont{Hamma et~al.}(2007)\citenamefont{Hamma, Miranda,
  Vasilevskiy, and Zorkani}}]{Hamma}
\bibinfo{author}{\bibfnamefont{M.}~\bibnamefont{Hamma}},
  \bibinfo{author}{\bibfnamefont{R.~P.} \bibnamefont{Miranda}},
  \bibinfo{author}{\bibfnamefont{M.~I.} \bibnamefont{Vasilevskiy}},
  \bibnamefont{and} \bibinfo{author}{\bibfnamefont{I.}~\bibnamefont{Zorkani}},
  \bibinfo{journal}{J. Phys.: Condensed Matter} \textbf{\bibinfo{volume}{19}},
  \bibinfo{pages}{346215} (\bibinfo{year}{2007}).

\bibitem[{\citenamefont{Chumanov et~al.}(1995)\citenamefont{Chumanov, Sokolov,
  Gregory, and Cotton}}]{Chumanov}
\bibinfo{author}{\bibfnamefont{G.}~\bibnamefont{Chumanov}},
  \bibinfo{author}{\bibfnamefont{K.}~\bibnamefont{Sokolov}},
  \bibinfo{author}{\bibfnamefont{B.~W.} \bibnamefont{Gregory}},
  \bibnamefont{and} \bibinfo{author}{\bibfnamefont{T.~M.}
  \bibnamefont{Cotton}}, \bibinfo{journal}{J. Phys. Chem.}
  \textbf{\bibinfo{volume}{99}}, \bibinfo{pages}{9466 } (\bibinfo{year}{1995}).

\bibitem[{Not({\natexlab{b}})}]{Note2}
\bibinfo{note}{Here we extend our consideration beyond the electrostatic
  approximation where the magnetic field was neglected.}

\bibitem[{\citenamefont{Born and Wolf}(1980)}]{Born-Wolf}
\bibinfo{author}{\bibfnamefont{M.}~\bibnamefont{Born}} \bibnamefont{and}
  \bibinfo{author}{\bibfnamefont{E.}~\bibnamefont{Wolf}},
  \emph{\bibinfo{title}{Principles of Optics}} (\bibinfo{publisher}{Pergamon
  Press, Oxford}, \bibinfo{year}{1980}).

\bibitem[{\citenamefont{Ebbesen et~al.}(1998)\citenamefont{Ebbesen, Lezec,
  Ghaemi, Thio, and Wolff}}]{Ebbesen98}
\bibinfo{author}{\bibfnamefont{T.~W.} \bibnamefont{Ebbesen}},
  \bibinfo{author}{\bibfnamefont{H.~J.} \bibnamefont{Lezec}},
  \bibinfo{author}{\bibfnamefont{H.~F.} \bibnamefont{Ghaemi}},
  \bibinfo{author}{\bibfnamefont{T.}~\bibnamefont{Thio}}, \bibnamefont{and}
  \bibinfo{author}{\bibfnamefont{P.~A.} \bibnamefont{Wolff}},
  \bibinfo{journal}{Nature (London)} \textbf{\bibinfo{volume}{391}}, \bibinfo{pages}{667
  } (\bibinfo{year}{1998}).

\bibitem[{\citenamefont{Alonso-González
  et~al.}(2014)\citenamefont{Alonso-González, Nikitin, Golmar, Centeno,
  Pesquera, Vélez, Chen, Navickaite, Koppens, Zurutuza
  et~al.}}]{Alonso-Gozalez2014}
\bibinfo{author}{\bibfnamefont{P.}~\bibnamefont{Alonso-Gonz\'alez}},
  \bibinfo{author}{\bibfnamefont{A.~Y.} \bibnamefont{Nikitin}},
  \bibinfo{author}{\bibfnamefont{F.}~\bibnamefont{Golmar}},
  \bibinfo{author}{\bibfnamefont{A.}~\bibnamefont{Centeno}},
  \bibinfo{author}{\bibfnamefont{A.}~\bibnamefont{Pesquera}},
  \bibinfo{author}{\bibfnamefont{S.}~\bibnamefont{V\'elez}},
  \bibinfo{author}{\bibfnamefont{J.}~\bibnamefont{Chen}},
  \bibinfo{author}{\bibfnamefont{G.}~\bibnamefont{Navickaite}},
  \bibinfo{author}{\bibfnamefont{F.}~\bibnamefont{Koppens}},
  \bibinfo{author}{\bibfnamefont{A.}~\bibnamefont{Zurutuza}},
  \bibnamefont{et~al.}, \bibinfo{journal}{Science}
  \textbf{\bibinfo{volume}{344}}, \bibinfo{pages}{1369 }
  (\bibinfo{year}{2014}).

\bibitem[{\citenamefont{Mishchenko}(2013)}]{Mishchenko}
\bibinfo{author}{\bibfnamefont{E.~G.} \bibnamefont{Mishchenko}},
  \bibinfo{journal}{Phys. Rev. B} \textbf{\bibinfo{volume}{88}},
  \bibinfo{pages}{115436} (\bibinfo{year}{2013}).

\bibitem[{\citenamefont{Echtermeyer et~al.}(2011)\citenamefont{Echtermeyer,
  Britnell, Jasnos, Lombardo, Gorbachev, Grigorenko, Geim, Ferrari, and
  Novoselov}}]{Echtermeyer}
\bibinfo{author}{\bibfnamefont{T.~J.} \bibnamefont{Echtermeyer}},
  \bibinfo{author}{\bibfnamefont{L.}~\bibnamefont{Britnell}},
  \bibinfo{author}{\bibfnamefont{P.~K.} \bibnamefont{Jasnos}},
  \bibinfo{author}{\bibfnamefont{A.}~\bibnamefont{Lombardo}},
  \bibinfo{author}{\bibfnamefont{R.~V.} \bibnamefont{Gorbachev}},
  \bibinfo{author}{\bibfnamefont{A.~N.} \bibnamefont{Grigorenko}},
  \bibinfo{author}{\bibfnamefont{A.~K.} \bibnamefont{Geim}},
  \bibinfo{author}{\bibfnamefont{A.~C.} \bibnamefont{Ferrari}},
  \bibnamefont{and} \bibinfo{author}{\bibfnamefont{K.~S.}
  \bibnamefont{Novoselov}}, \bibinfo{journal}{Nat. Commun.}
  \textbf{\bibinfo{volume}{2}}, \bibinfo{pages}{458} (\bibinfo{year}{2011}).

\bibitem[{\citenamefont{Pereira et~al.}(2013)\citenamefont{Pereira, Pereira,
  Smirnov, and Vasilevskiy}}]{EPL-Pereira}
\bibinfo{author}{\bibfnamefont{R.~M.} \bibnamefont{Pereira}},
  \bibinfo{author}{\bibfnamefont{P.}~\bibnamefont{Pereira}},
  \bibinfo{author}{\bibfnamefont{G.}~\bibnamefont{Smirnov}}, \bibnamefont{and}
  \bibinfo{author}{\bibfnamefont{M.~I.} \bibnamefont{Vasilevskiy}},
  \bibinfo{journal}{Europhys. Lett.} \textbf{\bibinfo{volume}{102}},
  \bibinfo{pages}{67001} (\bibinfo{year}{2013}).

\end{thebibliography}
\end{document}